\documentclass[journal]{IEEEtran}
\usepackage{bm}
\usepackage{array}
\usepackage{amsfonts}
\usepackage{balance}
\usepackage{booktabs}
\usepackage{amssymb}
\usepackage{amsmath}
\usepackage{tabularx}
\usepackage{subfigure}
\usepackage{graphicx}
\usepackage{url}
\usepackage{multirow}
\usepackage{rotating}
\usepackage{algorithm}
\usepackage{algorithmic}
\usepackage{footnote}
\usepackage{threeparttable}
\usepackage{cleveref}
\usepackage{mathtools}
\usepackage{tabularx}
\usepackage{float}
\usepackage{cite}
\usepackage{epstopdf}
\usepackage{ntheorem}

\crefname{equation}{Eq.}{Eqs.}
\crefname{figure}{Fig.}{Figs.}

\newcolumntype{Y}{>{\centering\arraybackslash}X}
\begin{document}
\title{G$^3$SR: Global Graph Guided Session-based Recommendation}

\author{Zhi-Hong~Deng,
        Chang-Dong~Wang,~\IEEEmembership{Member,~IEEE,~}% <-this % stops a space
        Ling Huang,~\IEEEmembership{Member,~IEEE,~}
        \protect\\
        Jian-Huang Lai,~\IEEEmembership{Senior Member,~IEEE} and
        Philip S. Yu,~\IEEEmembership{Fellow,~IEEE}
\thanks{This work was supported by NSFC (61876193 and 62106079), Natural Science Foundation of Guangdong Province (2020A1515110337), Guangdong Province Key Laboratory of Computational Science at the Sun Yat-sen University (2020B1212060032), Open Foundation of Guangdong Provincial Key Laboratory of Public Finance and Taxation with Big Data Application, and NSF under grants III-1763325, III-1909323, III-2106758, and SaTC-1930941.}
\thanks{Z.-H. Deng and C.-D. Wang are with the School of Computer Science and Engineering, Sun Yat-sen University, Guangzhou, China, Guangdong Province Key Laboratory of Computational Science, Guangzhou, China, and also with Key Laboratory of Machine Intelligence and Advanced Computing, Ministry of Education, China.
E-mail: family\_ld@foxmail.com, changdongwang@hotmail.com.}% <-this % stops a space
\thanks{Jian-Huang Lai is with School of Computer Science and Engineering, Sun Yat-sen
University, Guangzhou, China, Guangdong Key Laboratory of Information Security Technology, Guangzhou, China,  and Key Laboratory of Machine Intelligence and Advanced Computing, Ministry of Education, China.
E-mail: stsljh@mail.sysu.edu.cn.}
\thanks{Ling Huang is with the College of Mathematics and Informatics, South China Agricultural University, Guangzhou, China, and Guangdong Provincial Key Laboratory of Public Finance and Taxation with Big Data Application.
E-mail: huanglinghl@hotmail.com.
}
\thanks{P. S. Yu is with University of Illinois at Chicago, Chicago, IL, U.S.A.
E-mail: psyu@cs.uic.edu.}% <-this % stops a space
\thanks{Corresponding author: Chang-Dong Wang.}
}

\markboth{IEEE Transactions on Neural Networks and Learning Systems}%
{\MakeLowercase{\textit{Deng et al.}}: G$^3$SR}

\maketitle

\begin{abstract}
  Session-based recommendation tries to make use of anonymous session data to deliver high-quality recommendation under the condition that user-profiles and the complete historical behavioral data of a target user are unavailable. Previous works consider each session individually and try to capture user interests within a session. Despite their encouraging results, these models can only perceive intra-session items and cannot draw upon the massive historical relational information. To solve this problem, we propose a novel method named G$^3$SR (Global Graph Guided Session-based Recommendation). G$^3$SR decomposes the session-based recommendation workflow into two steps. First, a global graph is built upon all session data, from which the global item representations are learned in an unsupervised manner. Then, these representations are refined on session graphs under the graph networks, and a readout function is used to generate session representations for each session. Extensive experiments on two real-world benchmark datasets show remarkable and consistent improvements of the G$^3$SR method over the state-of-the-art methods, especially for cold items.
\end{abstract}

\IEEEpeerreviewmaketitle

\section{Introduction}
\label{sec:introduction}
\IEEEPARstart{W}{ith} the population of the Internet and rich internet services, recommender systems have become a widely used fundamental technique~\cite{DBLP:journals/tkde/MaKL12}. On the one hand, recommender systems help users alleviate information overload problems by finding the information they need rapidly and accurately, which significantly improves the user experience. On the other hand, high quality recommendation can lead to more desirable services and therefore facilitate greater retention of users. Traditional recommender systems provide personalized recommendations based on users' historical behaviors~\cite{cheng2019mmalfm,DBLP:journals/tnn/WangNL20,DBLP:journals/tnn/WangZYXD21}. However, users' identities may be unknown in some scenarios, i.e., unregistered users may want to experience the provided services to decide whether or not to use it, while some users may refuse to log in for privacy concerns. In such situations, it is still important to provide accurate recommendations, but the system can only observe the current session instead of adequate historical user-item interactions. This is the so-called session-based recommendation that has drawn an increasing amount of attention in the field of recommender systems in recent years.

Early efforts have been made in studying sequential recommendation problems such as next-basket prediction with Markov chains methods~\cite{shani2005mdp,rendle2010factorizing,DBLP:journals/tkde/HuangHSLZCS15,DBLP:journals/tkde/YeXGD19}. However, these methods are based on complete historical user data and have a strong independence assumption that the user's next behavior only relies on the current one, which significantly limits the performance. In 2015, Hidasi et al. first propose the session-based recommendation problem and devise a method called GRU4Rec~\cite{hidasi2016session} based on Recurrent Neural Networks (RNNs) to solve it. Since then, RNNs have become a popular choice for solving session-based recommendation and offer remarkable results. For example, Li et al. propose {the Neural Attentive Recommendation Machine (NARM)~\cite{li2017neural}}, a model that tries to capture sequential behavioral features as well as users' main purposes by using a two-way RNN model architecture. {Short-term Attention/Memory Priority Model (STAMP)~\cite{liu2018stamp}} replaces RNNs with Multilayer Perceptron (MLP) enhanced by the attention mechanism, which achieves impressive results. More recently, Wu et al. propose SR-GNN~\cite{wu2019session} which is able to capture higher-order item transitions by introducing Graph Neural Networks (GNN).

\begin{figure}[!t]
    \centering
    \includegraphics[width=0.9\linewidth]{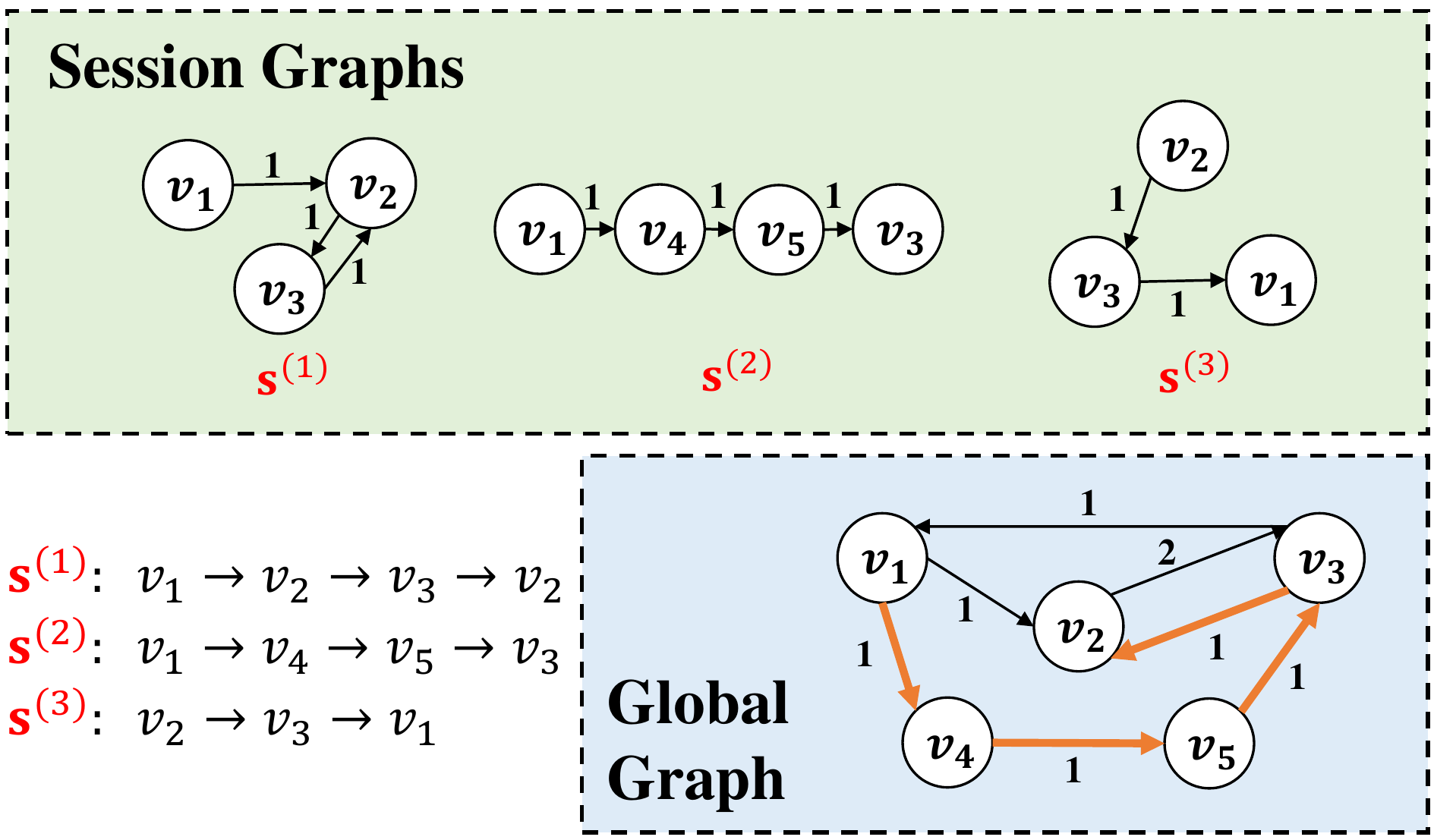}
    \caption{An example of session graphs and a global graph built on the same session data.}
    \label{fig:example_graph}
\end{figure}

Despite their promising results, all the above methods and models are trained on session graphs {which only model intra-session item-to-item relations, and fail to make full use of global information}. Figure~\ref{fig:example_graph} shows a simple example of session graphs and the global graph built on the same session data. In this example, sessions are denoted as $\mathbf{s}^{(i)}$ and items are denoted as $v_j$. Based on the given session sequences, three session graphs and one global graph are built, where an edge $v_i \rightarrow v_j$ is added to the graph if $v_i$ and $v_j$ are observed to occur consecutively. The existing methods try to capture information based on session graphs. Although session graphs can reflect part of the complex item-to-item transitions, {the global information implied in the global graph can be very useful}. For example, if the system is requested to recommend an item for session $\mathbf{s}^{(2)}$, $v_2$ is a reasonable choice since the user hasn't viewed it yet and it often appears along with $v_1$ and $v_3$. Explicit relations $v_1 \rightarrow v_2$ and $v_3 \rightarrow v_2$ can be observed directly in the global graph while the model can only learn implicit relations from data if it relies solely on session graphs. Since $v_2$ has never appeared together with $v_4$ and $v_5$, it may not be recommended in the latter case. Although intuitively, a model can implicitly learn global information with a sufficient amount of data, {the extremely sparse nature of the recommendation scenario makes it difficult. In practice, the problem revealed by the above example is ubiquitous since short sessions dominate the data set. Instead of striving to learn global information from tens of thousands of short sessions, constructing a global graph to aggregate the information implied in the overall data is a promising solution.} Still, learning directly from the global graph is highly challenging since it introduces additional relational information together with noises that are irrelevant to the current session and this can severely downgrade the performance as shown in~\cite{wu2019session}.

Besides, items in a session graph should contribute differently to the session representation since user's interest drifts with time. Previous works~\cite{li2017neural, liu2018stamp} have revealed that not all items in the current session are relevant to the next click and the last few items are usually the most crucial ones. The majority of the existing methods adopt the attention mechanism to learn importance weights for different items in the current session. Although these methods have achieved fascinating results, the attention mechanism is purely data-driven and computationally expensive. We believe that incorporating human knowledge (``prior'') is at least equally important as learning from the data. In particular, an empirically effective prior for session-based recommendations is that the more recently clicked items are of higher significance.

To tackle the above problems, we propose a novel method named Global Graph Guided Session-based Recommendation (G$^3$SR). {Inspired by the recent progress of pre-training\cite{goyal2017accurate, radford2018improving, schneider2019wav2vec}}, we introduce an unsupervised pre-training process to learn global item-to-item relations. First, the global graph is constructed on all session data. Second, graph embedding techniques are applied to learn the global representation for each item. We then use a graph network to refine these item representations based on session graphs. Unlike previous works, we devise a parameter-free graph readout mechanism to obtain the final session representation.

The main contributions of this paper are as follows.
\begin{itemize}
\item We introduce an unsupervised pre-training process to extract global item-to-item relational information. On the one hand, it provides richer semantic item representations for the downstream session-based recommendation task. On the other hand, it improves performance while reducing the amount of data required for the supervised training process. It offers a new perspective to solve the session-based recommendation problem.
\item We propose a parameter-free readout mechanism which uses the exponential decaying strategy to aggregate item representations into a session representation. This mechanism is highly competitive with the attention mechanism that is widely used in previous works, which sheds light on potential problems in current mainstream attention-based methods and calls for improved scientific practices.
\item Extensive experiments have been conducted on real-world datasets to demonstrate the effectiveness of G$^3$SR. We also conduct abundant ablation studies and sensitivity analyses which provide practical insights into the nature of the session-based recommendation problem. Our experimental results also show that the proposed method outperforms the state-of-the-art methods on cold items.
\end{itemize}

The rest of this paper is organized as follows. We first review some related literatures in Section~\ref{sec:relatedwork}. Then we describe the proposed method in Section~\ref{sec:methods}. Experiment results and analysis are demonstrated in Section~\ref{sec:experiments}. Finally, we conclude the paper in Section~\ref{sec:conclusion}.

\section{Related Work}
\label{sec:relatedwork}
In this section, we first review traditional methods for session-based recommendation, in comparison with deep learning-based methods. We then introduce the pre-training technique and finish this section by discussing the difference between the proposed method and previous methods.

\textbf{Traditional methods}. For the general recommendation problem, Matrix Factorization (MF)~\cite{koren2009matrix, he2017neural, deng2019deepcf,DBLP:journals/tnn/HeTDHRC20} is one of the most popular solutions. MF aims to decompose the user-item rating matrix and assumes that users and items can, therefore, be mapped into a common latent space where they can be directly compared. This method doesn't fit the session-based recommendation problem well since session data are anonymous and the rating matrix can be extremely sparse if we treat each session as a user. Item-based Collaborative Filtering (IBCF)~\cite{sarwar2001item, linden2003amazon, davidson2010youtube} is an alternative since it only relies on item similarities which can be calculated based on the co-occurrence in the session data. However, IBCF fails to capture the sequential relations because it recommends the next item merely based on the last one.

To model sequential data, some Markov chains (MC) based methods have been developed~\cite{zimdars2001using, shani2005mdp, rendle2010factorizing}. Shani et al. use Markov Decision Processes (MDPs) to solve the problem~\cite{shani2005mdp}, which can generate recommendations based on the learned transition probability matrix. Rendle et al. propose the {Factorizing Personalized Markov Chains model (FPMC)}~\cite{rendle2010factorizing} that combines the power of MF and MC to simultaneously model users' general interests and sequential behaviour patterns. Nevertheless, these methods suffer a huge drawback that the assumption on conditional independence is too strong.

\textbf{Deep learning-based methods}. With the rise of deep learning, an increasing number of research works begin to use deep learning techniques to solve recommendation problems~\cite{cui20MVRNN,Zhao20where,Zhang:TNNLS:2021}. Hidasi et al. first define the session-based recommendation problem which aims to recommend the next item based on an anonymous session and they propose the GRU4Rec model~\cite{hidasi2016session} based on recurrent neural networks. Based on GRU4Rec, Tan et al. further enhance the performance by employing a few useful techniques such as data augmentation and knowledge distillation~\cite{tan2016improved}. To better manage sequential behaviours and user's main purpose simultaneously, Li et al. propose {NARM}~\cite{li2017neural}, which uses the encoder-decoder architecture enhanced by the attention mechanism. Later, Liu et al. propose {STAMP}~\cite{liu2018stamp} based on simple MLP network, which is capable of capturing users' long-term interests as well as their short-term attention.

{A}s graph neural networks have achieved success in many different tasks, they have also been used in solving different recommendation problems~{\cite{zhao2019intentgc,liu2020a2,yang2021interpretable,liu2021interest}}. Wu et al. first employ graph neural networks to solve the session-based recommendation problem. They propose SR-GNN~\cite{wu2019session} which models session sequences as graph-structured data and use a graph neural network to capture complex item transitions. Xu et al. further propose a graph contextualized self-attention model (GC-SAN)~\cite{xu2019graph}, which enhances the graph neural network with a self-attention readout module. Qiu et al. improve the graph neural network by considering the inherent order of item transitions, where a weight graph attention network is used to learn item embeddings for each item in a session~\cite{qiu2019rethinking}{. More recently, there are two contemporary research works that also study the usage of global {information~\cite{qiu2020,wang2020}}. Qiu et al. propose the Full Graph Neural Network (FGNN)~\cite{qiu2020} along with a concept named Broadly Connected Session (BCS) graph which falls in between the session graph and the global graph. They use a Weighted Graph Attention Network (WGAT) to learn on this BCS graph rather than the session graph. Wang et al. propose the Global Context Enhanced Graph Neural Networks (GCE-GNN)~\cite{wang2020} which decomposes the representation learning process into two pathways. One is done on the session graph and the other one is done on the global graph. The representations are then added together and further enhanced by combining positional information. Note that these works are orthogonal to our research, since they both focus on designing more sophisticated model architectures and attention methods to implement an end-to-end recommendation algorithm, while we introduce a {pre-train}ing process and strive to use it to simplify the downstream recommendation problem.}

\textbf{Pre-training}. In general, most of the supervised tasks in Computer Vision (CV), Natural Language Processing (NLP) and Speech Recognition (SR) require huge amounts of labelled data to learn state-of-the-art models, but obtaining such data can be extremely expensive and sometimes impossible. Thus, pre-training~\cite{goyal2017accurate, radford2018improving, schneider2019wav2vec} has appeared as an important and effective technique in recent years because representations learned with available substantial amounts of labelled or unlabeled data can benefit various downstream tasks where only limited amounts of data are available. {For example, in NLP}, pre-trained language models~\cite{peters2018deep, devlin2018bert} have made significant breakthroughs in the past {few} years in text classification~\cite{lample2019cross}, machine translation~\cite{edunov2019pre}, etc. {While the target of pre-training and fine-tuning may be completely different, it allows to make better use of available data in terms of learning useful representations. Besides, unsupervised pre-training is empirically proved to have better generalization and serves as a regularization method~\cite{erhan2010does}}. More recently, pre-training graph neural networks~\cite{hu2019pre} is proposed to improve performance on tasks operated on graph-structured data.

Although an increasing amount of research focuses on the session-based recommendation problem, unlike previous studies, we take the global item-to-item relations into consideration and explore how it can further improve the performance as well as reduce the amount of data required in the training process. We achieve this goal by introducing an unsupervised pre-training process to learn initialized representations for each item. Besides, we also investigate the readout function for the graph network. Our results point out that parameter-free methods can achieve highly competitive performance compared with data-driven attention-based methods.

\section{The Proposed Method}
\label{sec:methods}

\subsection{Problem Formulation}
\label{subsec:problemFormulation}

A session-based recommender system aims to provide attractive recommendations for anonymous users given current sessions. As a result, it requires the model to accurately capture users' interests by making use of short sessions instead of their complete historical interaction records.

Suppose there are $m$ items and $n$ sessions in total. Let $\mathcal{I}=\{v_1, v_2, \cdots, v_m\}$ and $\mathcal{D} = \{\mathbf{s}^{(1)}, \mathbf{s}^{(2)}, \cdots, \mathbf{s}^{(n)}\}$ denote the item set and the anonymous session data respectively with $\mathbf{s}^{(i)}$ representing the $i$-th session. Each session $\mathbf{s}^{(i)}$ is an ordered sequence $[v_1^{(i)}, v_2^{(i)}, \cdots, v_{|\mathbf{s}^{(i)}|}^{(i)}]$, where $v_j^{(i)} \in \mathcal{I}$ is the $j$-th clicked item of the $i$-th session. Our goal is to predict the next item $v_{|\mathbf{s}^{(i)}|+1}^{(i)}$ for any session $\mathbf{s}^{(i)}$.

\subsection{The Unsupervised Pre-training Process}
\label{subsec:unsupervisedPretraining}
Previous studies ignore global item information and focus on intra-session learning. Here we introduce an unsupervised pre-training process to learn such global information and argue that they are beneficial to the subsequent recommendation process. We start by constructing the global graph and generating random walks. The item representations are then learnt by maximizing the probability of predicting contextual items.

\subsubsection{Constructing the global graph}
\label{subsubsec:constructGlobalGraph}
The global graph $\mathcal{G}=(\mathcal{V}, \mathcal{E}, \mathbf{W})$ is constructed based upon all session data~(denoted as $\mathcal{D}$). In this case, the node set $\mathcal{V}$ is equivalent to the full item set $\mathcal{I}$, $\mathcal{E}$ denotes all the edges in $\mathcal{G}$, and $\mathbf{W}$ denotes the weight matrix of each node pair. If two items $v_i$ and $v_j$ are observed to occur consecutively in $\mathcal{D}$, they are linked together with a directed edge $e_{i, j}$ and the weight of this edge is the corresponding co-occurence frequency in $\mathcal{D}$. For example, in  \figurename~\ref{fig:example_graph}, $v_2$ and $v_3$ occur consecutively twice (in $\mathbf{s}^{(1)}$ and $\mathbf{s}^{(3)}$ respectively), so the weight of $e_{2, 3}$ is 2.

\begin{figure*}[!t]
    \centering
    \includegraphics[width=0.95\linewidth]{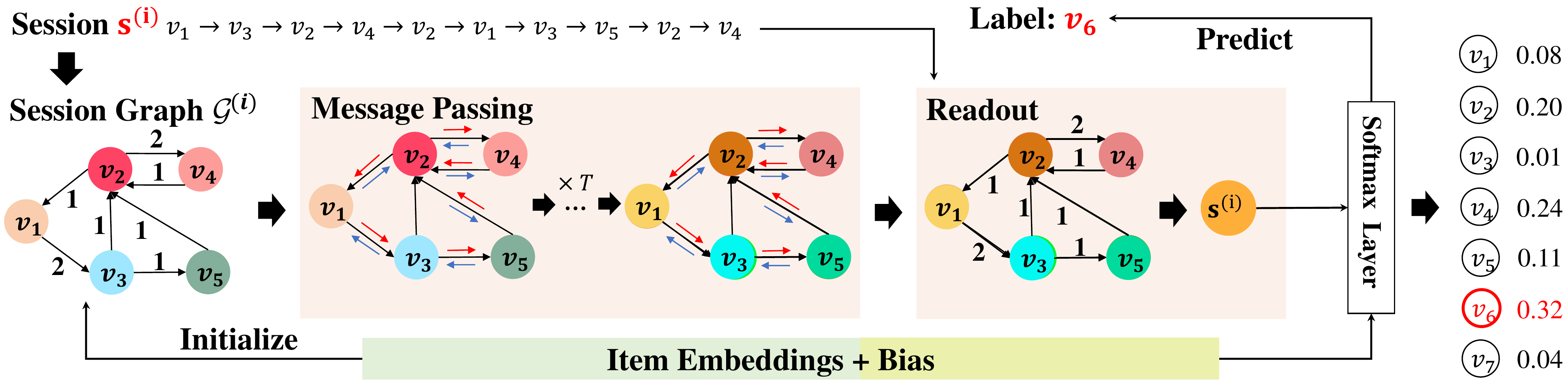}
    \caption{The workflow of the supervised learning process.}
    \label{fig:architecture}
\end{figure*}

\subsubsection{Learning item embeddings}
\label{subsubsec:learn_embedding}
With the weighted directed graph $\mathcal{G}$, we use a weighted version of \emph{node2vec}~\cite{grover2016node2vec} to learn embeddings for different items. The basic assumption behind it is that adjacent nodes should have similar representations. To be specific, we first generate a set of random walks on $\mathcal{G}$ according to the weight matrix $\mathbf{W}$. Suppose the previous transition in the current walk is $v_i \rightarrow v_s$, the transition probability between the source node $v_s$ and any possible target node $v_t$ is defined as
\begin{equation}
P\left(v_{t} | v_{s}\right)=
\left\{\begin{array}{ll}
{\frac{\alpha(v_{i}, v_{t})\mathbf{W}_{s, t}}{\sum_{j \in \mathcal{N}\left(v_{s}\right)} \alpha(v_{i}, v_{j})\mathbf{W}_{s, j}},}
& {v_{t} \in \mathcal{N}\left(v_{s}\right)} \\
{0,}
& {\text{otherwise,}}
\end{array}\right.
\end{equation}
where $\mathcal{N}(v_{s})$ is the set of neighborhood nodes for $v_{s}$ and $\alpha(v_{i}, v_{j})$ is an adjustment factor that corrects the weight $\mathbf{W}_{s, j}$ between the source node $v_s$ and a potential target node $v_j$. $\alpha(v_{i}, v_{j})$ is defined as
\begin{equation}
\alpha(v_{i}, v_{j})=
\left\{\begin{array}{ll}
{\frac{1}{p}} & {\text { if } d_{i, j}=0~(v_j=v_i)}\\
{1} & {\text { if } d_{i, j}=1~(v_j\in\mathcal{N}(v_s) \land v_j\in\mathcal{N}(v_i))}\\
{\frac{1}{q}} & {\text { if } d_{i, j}=2~(v_j\in\mathcal{N}(v_s) \land v_j\not\in\mathcal{N}(v_i)}
\end{array}\right.
\end{equation}
where $d_{i, j}\in\{0, 1, 2\}$ denotes the distance from $v_i$ to $v_j$. $d_{i, j}=0$ means the current walk repeats the previous transition in a reverse way, i.e., $v_s \rightarrow v_i$. $d_{i, j}=1$ means the current walk proceeds with a target node within the common neighbors of $v_i$ and $v_s$. And $d_{i, j}=2$ means the current walk procceeds with $v_s$'s neighbor node that is not adjacent to $v_i$. The probabilities of these three types of transitions are controled by the \emph{return parameter} $p\in \mathbb{R}$ and the \emph{in-out parameter} $q \in \mathbb{R}$. If the return parameter $p$ is set to a large value ($>\max (q, 1)$), the current walk would be less likely to return to the previous node $v_i$. Otherwise, it would encourage the walk to continue with node $v_i$, i.e., $v_t=v_i$. If the in-out parameter $q$ is set to a large value ($q>1$), the current walk would be always biased toward nodes near the previous node $v_i$ and can therefore better exploit the local structure; otherwise, it would encourage the walk to explore nodes that are further away from $v_i$. Note that, $d_{i, j}$ must be one of $\{0, 1, 2\}$, otherwise the node $v_j$ can not be reached from the source node $v_s$, i.e., $v_j \not\in \mathcal{N}(v_s)$.

After collecting a fixed number of walks for each {node}, item (node) embeddings are then learned via the Skip-Gram algorithm which maximizes the probability of observing the context node of $v_i$ given its embedding $\phi(v_i)$:
\begin{equation}
\label{preObjective}
\max_\phi \sum_{v_i \in \mathcal{V}} \log \text{Pr}(\{v_{i-c}, \cdots, v_{i+c} \backslash v_i\}|\phi(v_i)),
\end{equation}
where $c$ is the size of context window. Using the conditional independence assumption, this optimization problem can be reduced to:
\begin{equation}
\label{objective}
\max_\phi \sum_{v_i \in \mathcal{V}} \sum_{j=i-c, j \neq i}^{i+c} \log \text{Pr}(v_j|\phi(v_i)),
\end{equation}
where the conditional probability $\text{Pr}(v_j|\phi(v_i))$ is modeled by a softmax unit that takes the dot product of $\phi(v_j)$ and $\phi(v_i)$ as input:
\begin{equation}
\text{Pr}(v_j|\phi(v_i)) = \frac{\exp (\phi(v_j)^{T}\phi(v_i))}{\sum_{v_k \in \mathcal{V}}\exp (\phi(v_k)^{T}\phi(v_i))}
\end{equation}

Since the denominator of $\text{Pr}(v_j|\phi(v_i))$ is computationally expensive, the optimization problem (\ref{objective}) is usually solved by adopting negative sampling. As a result, the final objective function is
\begin{equation}
\small
\label{finalObjective}
\min_\phi \sum_{v_i \in \mathcal{V}} \sum_{j=i-c, j \neq i}^{i+c} \log \sigma(\phi(v_j)^{T}\phi(v_i)) + \sum_{v_k \in \mathcal{S}(v_i)}\log \sigma(-\phi(v_k)^{T}\phi(v_i)),
\end{equation}
where $\mathcal{S}(v_i)$ is a set of negative samples for node $v_i$. Finally, item embeddings $\phi(v_i) \in \mathbb{R}^d$ are learnt by optimizing problem (\ref{finalObjective}) with stochastic gradient descent.

Using random walks instead of session sequences to learn item embeddings has several benefits. First, it assures all nodes are eligible to be the origin and will generate a fixed number of sequences, which contributes to learn good semantic representations for cold items and can alleviate the long tail phenomenon widely existed in recommendation. Some experiments are reported in Section~\ref{subsec:case} to demonstrate the effect. Second, it explores dependency relations that do not occurred in session data. Last but not least, by introducing the return parameter $p$ and the in-out parameter $q$, we can generate sequences flexibly according to the properties we want.

\begin{figure}[!t]
    \centering
    \includegraphics[width=1.0\linewidth]{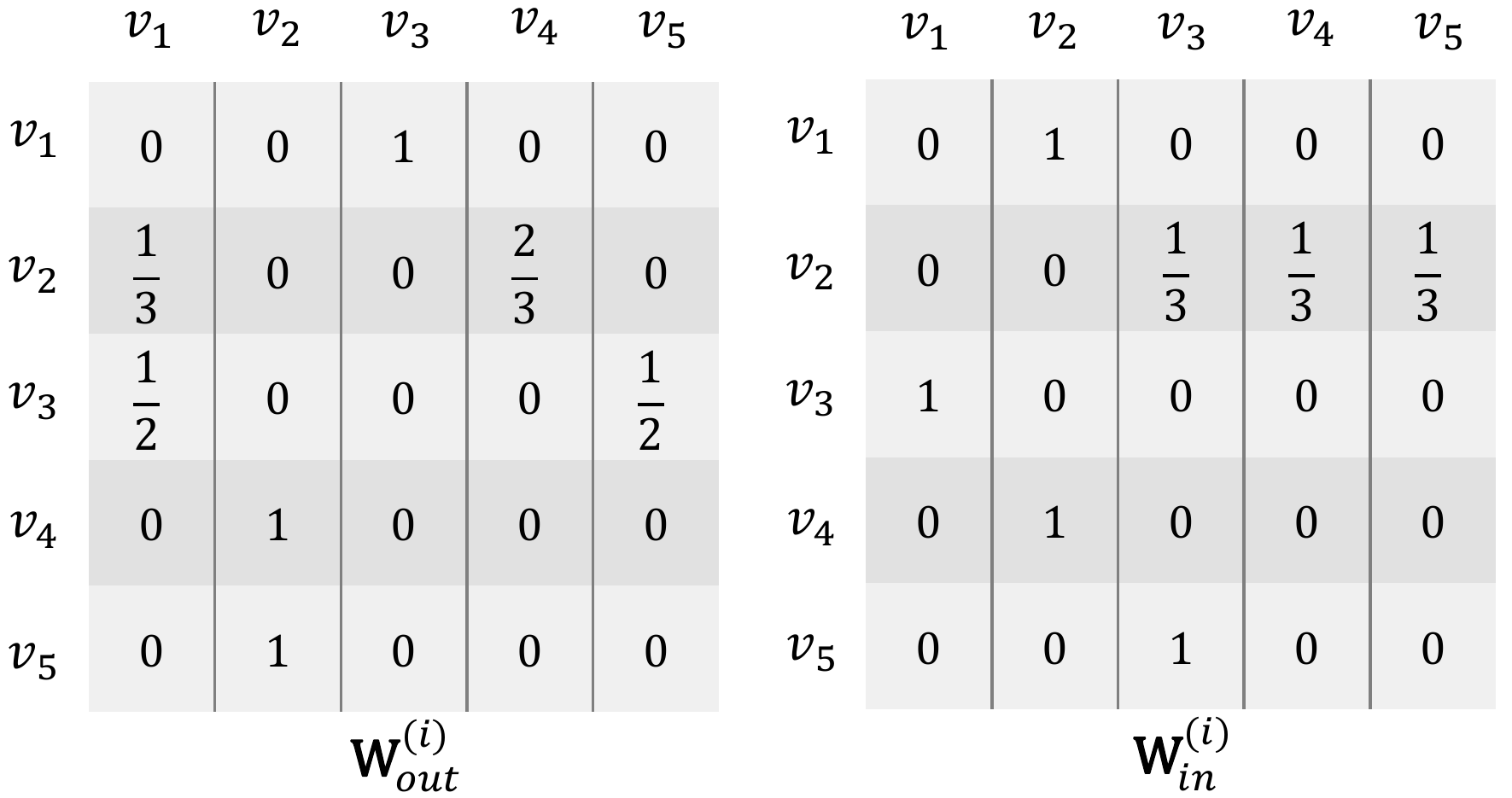}
    \caption{The normalized weight matrices for the session graph $\mathcal{G}^{(i)}$ presented in \figurename~\ref{fig:architecture}.}
    \label{fig:weightMatrix}
\end{figure}

\subsection{The Supervised Learning Process}
The overall workflow of the supervised learning process is shown in \figurename~\ref{fig:architecture}. A session sequence $\mathbf{s}^{(i)}$ is first turned into a session graph $\mathcal{G}^{(i)}$ with nodes' features initialized by item embeddings pre-trained in the proposed unsupervised pre-training process. Note that we propose to introduce a task-specific bias for each item to overcome the gap between the two processes. We then use a graph network to combine and update node representations. Last but not least, unlike other methods, the session representation is aggregated via an exponential decaying readout function and then used to calculate the output distribution.

\label{subsec:supervisedLearning}
\subsubsection{Constructing session graphs}
Unlike the global graph, a session graph $\mathcal{G}^{(i)}=(\mathcal{V}^{(i)}, \mathcal{E}^{(i)}, \mathbf{W}^{(i)})$ is built on a single session $\mathbf{s}^{(i)}$. The node set $\mathcal{V}^{(i)}$ and the edge set $\mathcal{E}^{(i)}$ are defined in the same way as in Section~\ref{subsubsec:constructGlobalGraph} but with respect to the session graph $\mathbf{s}^{(i)}$. Note that $\mathbf{W}^{(i)} = [\mathbf{W}^{(i)}_{out};\mathbf{W}^{(i)}_{in}]$ where $\mathbf{W}^{(i)}_{out} \in \mathbb{R}^{|\mathcal{V}^{(i)}| \times |\mathcal{V}^{(i)}|}$ and $\mathbf{W}^{(i)}_{in} \in \mathbb{R}^{|\mathcal{V}^{(i)}| \times |\mathcal{V}^{(i)}|}$ are the normalized weight matrices for the outgoing edges and incoming edges respectively. In this way, messages can flow bi-directionally in the subsequent message passing process, which is beneficial to enrich the information that a node receives. An example of $\mathbf{W}^{(i)}$ is demonstrated in \figurename~\ref{fig:weightMatrix}.

\subsubsection{Setting up node representations}
\label{subsubsec:initialization}

For each item (node) $v$ in a session graph $\mathcal{G}^{(i)}$, the representation contains two parts, i.e., $\mathbf{h}_v^{<0>}=\phi(v)+\beta(v)$. The first part $\phi(v)$ is a fixed vector learned in the unsupervised pre-training process, which is global-graph-aware and has abundant semantic information. The second part $\beta(v)$ is a learnable vector with random initialization. This vector is referred to as a bias term in this paper since it biases $\phi(v)$ towards an item representation that fits the downstream session-based recommendation task. We introduce this term for two reasons. First, transferring the knowledge obtained from the pre-training process may introduce noises and hurt the performance of the target task unexpectedly, which is known as the ``negative transfer'' phenomenon~\cite{pan2010a} in transfer learning. The bias term can learn to offset such harmful noises. Second, since the global information is not enough to accurately describe a session, the bias term serves as a residual term that aims to complement the information needed to solve a specific recommendation task.

\subsubsection{Message passing}

To combine and update item information, we use a message passing graph network to further process the session graphs. Specifically, in each step, we use the Gated Graph Neural Network (GGNN)~\cite{li2015gated} to update the representation of each node. From \figurename~\ref{fig:architecture} we can see that a node receives messages from both of its incoming edges (red arrows) and outgoing edges (blue arrows) to update itself during this process. The message that a node $v$ receives at time step $t$ is defined as
\begin{equation}
\label{message}
\mathbf{m}_v^{<t>} = \mathbf{W}_{v:}[\mathbf{h}_1^{<t-1>}, \mathbf{h}_2^{<t-1>}, \cdots, \mathbf{h}_n^{<t-1>}]^\mathrm{T} + \mathbf{b},
\end{equation}
where $\mathbf{W}_{v:} \in \mathbb{R}^{2|\mathcal{V}|}$ is the row in $\mathbf{W}$ corresponding to node $v$, $\mathbf{h}_v^{<t-1>}$ stands for the representation of node $v$ at the previous time step, and $\mathbf{b} \in \mathbb{R}^{2d}$ is a bias vector. The message $\mathbf{m}_v^{<t>}$ is then used to update node $v$'s representation in the following manner:
\begin{equation}
\label{messagePassing}
\begin{split}
\mathbf{z}_v^{<t>} &= \sigma(\mathbf{P}^z \mathbf{m}_v^{<t>} + \mathbf{Q}^z \mathbf{h}_v^{<t-1>})\\
\mathbf{r}_v^{<t>} &= \sigma(\mathbf{P}^r \mathbf{m}_v^{<t>} + \mathbf{Q}^r \mathbf{h}_v^{<t-1>})\\
\widetilde{\mathbf{h}_v^{<t>}} &= \text{tanh}(\mathbf{P} \mathbf{m}_v^{<t>} + \mathbf{Q} (\mathbf{r}_v^{<t>} \odot \mathbf{h}_v^{<t-1>}))\\
\mathbf{h}_v^{<t>} &= (1-\mathbf{z}_v^{<t>}) \odot \mathbf{h}_v^{<t-1>} + \mathbf{z}_v^{<t>} \odot \widetilde{\mathbf{h}_v^{<t>}},
\end{split}
\end{equation}
where $\mathbf{P}^z, \mathbf{P}^r, \mathbf{P} \in \mathbb{R}^{2d \times d}$, $\mathbf{Q}^z, \mathbf{Q}^r, \mathbf{Q} \in \mathbb{R}^{d \times d}$ are learnable parameter matrices. $\mathbf{z}_v$ and $\mathbf{r}_v$ are the update and reset gates that decide the quantity of information to be saved or discarded. If we stack $T$ layers of graph networks, the information will be passed over $T$ time steps and therefore a node can receive messages sent by its $T$-hop neighbors. Note that the above formulas are discussed under a session graph $\mathcal{G}^{(i)}$. We ignore the superscript $(i)$ to keep them simple and neat, similarly hereinafter.

\subsubsection{Readout and make recommendations}
\label{subsubsec:readout}
Suppose we stack $T$ layers of graph networks, the message passing process outputs a representation $\mathbf{h}_v^{<T>}$ for each node $v$. As shown in \figurename~\ref{fig:architecture}, the next step is to readout the session graph $\mathcal{G}^{(i)}$ to obtain a single representation for the whole session sequence $\mathbf{s}^{(i)}$. To reach this goal, previous works such as SR-GNN~\cite{wu2019session} adopt the short-term attention priority mechanism and some common graph neural networks use permutation invariant function such as mean and summation~\cite{xu2019how}. Since user's interest drifts along with time, the nature of the session-based recommendation problem is highly sensitive to the order that an item appears. In this paper, we adopt an exponential decaying strategy to aggregate item (node) representations into a session representation:
\begin{align}
\label{eq:expReadout}
\begin{split}
\mathbf{h}_{\mathbf{s}^{(i)}} &= \sum_{v \in \mathbf{s}^{(i)}} w_v \mathbf{h}_v^{<T>}\text{, where}\\
w_v &= \frac{\exp (-(|\mathbf{s}^{(i)}|-\text{pos}_{\mathbf{s}^{(i)}}(v))/\tau)}{\sum_{u \in \mathbf{s}^{(i)}} \exp (-(|\mathbf{s}^{(i)}|-\text{pos}_{\mathbf{s}^{(i)}}(u))/\tau)}
\end{split}
\end{align}
$\text{pos}_{\mathbf{s}^{(i)}}(v)$ is the position of node $v$ in a session sequence $\mathbf{s}^{(i)}$. Since $|\mathbf{s}^{(i)}|$ is fixed for all items in $\mathbf{s}^{(i)}$, the later items have larger weights and therefore are more influential on the session representation $\mathbf{h}_{\mathbf{s}^{(i)}}$. The temperature parameter $\tau$ is a hyper-parameter controlling the degree of attenuation. If $\tau$ is set to a small value, the vast majority of the probability density will be accumulated on the last few clicks; otherwise, it decays softly along with time.

Finally, we use the session representation $\mathbf{h}_{\mathbf{s}^{(i)}}$ to predict the next item $v_{|\mathbf{s}^{(i)}|+1}^{(i)}$. Using a fully-connected neural network layer to compute the prediction can be problematic due to the gigantic output space for massive number of items. A viable option is to reuse the initial representation $\mathbf{h}_v^{<0>}$ for each node $v$. Specifically, the output distribution $\hat{\mathbf{y}} \in \mathbb{R}^{|\mathcal{I}|}$ is computed as:
\begin{equation}
\label{output}
\hat{\mathbf{y}} = \text{softmax}(\mathbf{h}_{\mathbf{s}^{(i)}}^{\mathrm{T}}\mathbf{h}_:^{<0>}),
\end{equation}
where $\mathbf{h}_:^{<0>} \in \mathbb{R}^{|\mathcal{I}| \times d}$ denotes the initial representations for all items. The ground-truth item can be converted to a one-hot vector $\mathbf{y} \in \mathbb{R}^{|\mathcal{I}|}$. Then the model parameters $\Theta$ is optimized by minimizing the categorical cross-entropy:
\begin{equation}
\label{CE}
\mathcal{L}(\Theta) = \sum_{i=1}^n \mathbf{y}^{(i)} \log(\hat{\mathbf{y}}^{(i)}) + (1-\mathbf{y}^{(i)}) \log(1-\hat{\mathbf{y}}^{(i)}).
\end{equation}

\section{Experiments and Analysis}
\label{sec:experiments}

In this section, we first describe the experimental setup including datasets, baselines, and evaluation metrics (Section~\ref{subsec:setup}). Then, we elaborate the experimental performance of the proposed G$^3$SR method compared with other methods. We start by discussing the comparison results (Section~\ref{subsec:comparison}) and then verify the effectiveness of G$^3$SR by performing detailed ablation studies (Section~\ref{subsec:ablation}). We also conduct sensitivity analyses of G$^3$SR under different hyper-parameters settings {(Section~\ref{subsec:sensitivity})}. Last but not least, we provide a case study on popular items and cold items to further illustrate the effect of the proposed method {(Section~\ref{subsec:case})}. To make our results fully reproducible, all the relevant source codes will be released upon acceptance.
\begin{table}[!t]
\centering
    \caption{Statistics of the datasets.}
    \label{statisticsOfDatasets}
    \footnotesize
    \centering
    \begin{tabular}{lccc@{}}
        \toprule
        Statistics&Yoochoose 1/64&Yoochoose 1/4&Diginetica\\
        \midrule
        \# of clicks&557,248&8,326,407&982,961\\
        \# of train sessions&369,859&5,917,745&719,470\\
        \# of test sessions&55,898&55,898&60,858\\
        \# of items&17,376&30,444&43,097\\
        Average length&5.11&4.70&4.13\\
        \bottomrule
    \end{tabular}
\end{table}
\begin{table}[!t]
\centering
    \caption{Default values of hyper-parameters for \emph{node2vec}.}
    \label{tab:node2vec}
    \footnotesize
    \centering
    \begin{tabular}{lc}
        \toprule
        Hyper-parameter&Default value\\
        \midrule
        embedding size&100\\
        walk length&80\\
        \# of walks per node&10\\
        window size&10\\
        return parameter $p$ & 0.25\\
        in-out parameter $q$ & 4\\
        \# of epochs&5\\
        \bottomrule
    \end{tabular}
\end{table}

\begin{table*}
\centering
\caption{Performance comparison between G$^3$SR and baseline methods over three benchmark datasets}
\label{tab:comp}
\begin{threeparttable}
\resizebox{\textwidth}{!}{
\begin{tabular}{lccccccccc}
\toprule
\multirow{2}{*}{Method} & \multicolumn{3}{c}{Yoochoose 1/64} & \multicolumn{3}{c}{Yoochoose 1/4} & \multicolumn{3}{c}{Diginetica} \\
&P@20    & MRR@20    &NDCG@20    & P@20    & MRR@20    &NDCG@20   &P@20    &MRR@20   &NDCG@20\\
\midrule
POP      & 6.71~(-88.93\%)  & 1.65~(-92.79\%)  & 2.92~(-90.71\%) & 1.33~(-97.77\%)  & 0.30~(-98.67\%)  & 0.55~(-98.22\%) & 0.89~(-96.98\%)  & 0.20~(-97.60\%)   & 0.36~(-97.27\%) \\
S-POP    & 30.44~(-49.80\%) & 18.35~(-19.83\%) & 20.23~(-35.63\%) & 27.08~(-54.51\%) & 17.75~(-21.46\%) & 19.78~(-36.07\%) & 21.06~(-28.49\%) & 13.68~(+64.23\%)   & 14.70~(+11.28\%) \\
Item-KNN & 51.60~(-14.91\%) & 21.81~(-4.72\%)  & 29.93~(-4.77\%) & 52.31~(-12.13\%) & 21.70~(-3.98\%)  & 29.68~(-4.07\%) & 35.75~(-21.39\%) & 11.57~(+38.90\%)   & 17.05~(+29.07\%) \\
BPR      & 31.31~(-48.37\%) & 12.08~(-47.23\%) & 17.89~(-43.08\%) & 3.40~(-94.29\%)  & 1.57~(-93.05\%)  & 2.46~( -92.05\%) & 5.24~(-82.21\%)  & 1.98~(-76.23\%)   & 3.22~(-75.62\%) \\
FPMC     & 45.62~(-24.77\%) & 15.01~(-34.43\%) & 19.68~(-37.38\%)             & -                & -                & - & 26.53~(-9.92\%)  & 6.95~(-16.57\%)   & 11.64~(-11.88\%) \\
\midrule
GRU4REC  & 60.64            & 22.89            & 31.43 & 59.53            & 22.60            & 30.94 & 29.45            & 8.33   & 13.21 \\
NARM     & 68.32~(+12.66\%) & 28.63~(+25.08\%) & 37.09~(+18.01\%) & 69.73~(+17.13\%) & 29.23~(+29.34\%) & 37.53~(+21.30\%) & 49.70~(+68.76\%) & 16.17~(+94.11\%)   & 23.52~(+78.05\%) \\
STAMP    & 68.74~(+13.36\%) & 29.67~(+29.62\%) & 38.41~(+22.21\%) & 70.44~(+18.33\%) & 30.00~(+32.74\%) & 38.85~(+25.57\%) & 45.64~(+54.97\%) & 14.32~(+71.91\%)   & 21.24~(+60.79\%) \\
SR-GNN   & 70.57~(+16.38\%) & \textbf{30.94}~(+35.17\%)    & 39.20~(+24.72\%)    & 71.36~(+19.87\%)    & \textbf{31.89}~(+41.11\%)    & \textbf{40.61}~(+31.25\%)    & 50.73~(+72.26\%)    & 17.59~(+111.16\%)   & 24.68~(+86.83\%) \\
\midrule
FGNN    & \textbf{71.75}~(+18.32\%) & \textbf{32.45}~(+15.77\%) & - & \textbf{72.48}
~(+21.75\%) & \textbf{32.71}~(+44.73\%) & - & 51.67~(+75.45\%) & 18.69~(+124.37\%) & - \\
GCE-GNN & \textbf{71.75}~(+18.32\%) & 30.43~(+32.94\%) & \textbf{39.92}~(+27.01\%) & 70.48~(+18.39\%) & 30.24~(+33.81\%) & 39.43~(+27.44\%) & \textbf{54.18}~(+83.97\%) & \textbf{19.01}~(+128.21\%) & \textbf{26.46}~(+100.30\%) \\
G$^3$SR  & \textbf{72.61}~(+19.74\%)    & 30.74~(+34.29\%)    & \textbf{40.29}~(28.19\%)    & \textbf{71.62}~(+20.31\%)    & 30.68~(+35.75\%)    & \textbf{40.02}~(29.35\%)    & \textbf{54.05}~(+83.53\%)    & \textbf{18.72}~(+124.73\%)   & \textbf{26.59}~(101.29\%) \\
\bottomrule
\end{tabular}}
\begin{tablenotes}
\item[*] \scriptsize Note that Yoochoose 1/64 and Yoochoose 1/4 share the same test set. While some methods perform well with more training data by using Yoochoose 1/4, the others work

    better on Yoochoose 1/64 since it is less noisy w.r.t. users' most recent interests. Most importantly, the time cost for training on Yoochoose 1/64 is far less than Yoochoose 1/4,

    i.e., 19+ mins/epoch v.s. 4+ hours/epoch. Therefore, for similar performance, the method works well on Yoochoose 1/64 is better.
\end{tablenotes}
\end{threeparttable}
\end{table*}
\subsection{Experimental Setup}
\label{subsec:setup}
\subsubsection{Datasets}
Following~\cite{li2017neural,liu2018stamp,wu2019session}, we use two publicly available benchmark datasets to evaluate the proposed method, {i.e., } \emph{Yoochoose}\footnote{\url{http://2015.recsyschallenge.com/challege.html}} and \emph{Diginetica}\footnote{\url{http://cikm2016.cs.iupui.edu/cikm-cup}}. Both of them are challenging datasets that contain massive session data. The former comes from the RecSys Challenge 2015 and the latter comes from the CIKM Cup 2016. Sessions of length 1 and items that appear less than 5 times are removed. The data augmentation technique proposed in~\cite{tan2016improved} is adopted to make better use of the data. The training/test sets splitting is based on time, i.e., the sessions of the last day and the last week are used to build the test sets for Yoochoose and Diginetica respectively. Note that, for the Yoochoose dataset, two smaller datasets are constructed with the most recent fractions 1/64 and 1/4 of session data in previous researches. This is because the original dataset is too large and directly training on this entire dataset yields poor results since behavioral patterns and user's interests change over time~\cite{li2017neural}. For both Yoochoose and Diginetica, the validation set is constructed using the most recent 10\% session data extracted from the training set. The statistics of datasets are summarized in Table~\ref{statisticsOfDatasets}.

\subsubsection{Baseline methods}
The roadmap of session-based recommendation starts from {GRU4REC~\cite{hidasi2016session}} which first proposes this problem and uses recurrent neural networks to process session sequences. {NARM~\cite{li2017neural} and STAMP~\cite{liu2018stamp}} are two milestones as they introduce well-designed attention mechanisms into this problem. Most recently, {SR-GNN~\cite{wu2019session}} adopts graph neural networks to model complex transitions between items and becomes the new {state-of-the-art}. In this paper, we further study the use of graph pre-training techniques that fits the session-based recommendation problem. Therefore, we consider three types of baseline methods, {namely, traditional methods, e.g. Item-KNN, Bayesian Personalized Ranking (BPR)~\cite{rendle2009bpr} and FPMC, deep-learning-based methods, e.g. GRU4REC, NARM and STAMP, and graph-aware methods, e.g. SR-GNN, FGNN and GCE-GNN:}
\begin{itemize}
\item \textbf{POP} and \textbf{S-POP} are two widely used {baselines} which recommend items based on item popularity in the training set and in the current session respectively.
\item \textbf{Item-KNN}~\cite{sarwar2001item} is an item-to-item collaborative filtering method, which recommends items similar to the last clicked item in the current session. Specifically, cosine similarity is used to measure the similarity.
\item \textbf{BPR}~\cite{rendle2009bpr} is a classical recommendation algorithm which optimizes the {MF} model with a pairwise ranking loss function via stochastic gradient descent.
\item \textbf{FPMC}~\cite{rendle2010factorizing} is a sequential prediction method based on Markov chain, used for next-basket recommendation. In order to make it work on session-based recommendation, the user latent representation is ignored when computing recommendation scores.
\item \textbf{GRU4REC}~\cite{hidasi2016session} is a deep learning model based on RNN. It utilizes session-parallel mini-batch training, mini-batch based output sampling and a well-designed TOP1 loss function to help the RNN model to fit into the session-based recommendation problem.
\item \textbf{NARM}~\cite{li2017neural} is a {method} which introduces attention mechanism to capture the user's main purpose. The main purpose is then combined with sequential behavioral feature and used as the final representation to generate the next item.
\item \textbf{STAMP}~\cite{liu2018stamp} is a {method} which utilizes simple MLP enhanced by attention mechanism to capture user's general interest and current interest of the current session simultaneously.
\item \textbf{SR-GNN}~\cite{wu2019session} is a state-of-the-art method that first introduces {GNN} to capture complex item transitions. To generate the next item of the current session, it uses the same idea as STAMP, i.e., utilizes an attention mechanism to capture user's general interest and current interest.
\item \textbf{FGNN}~\cite{qiu2020} and \textbf{GCE-GNN}~\cite{wang2020} are two contemporary approaches that also utilize global information in session-based recommendation but in different ways compared with G$^3$SR. FGNN extends the session graph to include global information and GCE-GNN modifies the model structure to introduce the information learned from the global graph. Both of these two approaches focus only on the supervision learning process.
\end{itemize}

\subsubsection{Evaluation measures}
Following the previous works, we use three widely adopted evaluation measures to evaluate the performance, namely Precision (P), Mean Reciprocal Rank (MRR), and Normalized Discounted Cumulative Gain (NDCG). In particular, these metrics are calculated on the top-K recommendations. P@$K$ measures the recommendation accuracy, i.e., whether the ground-truth item shows up among the top-K results or not. On the other hand, MRR@$K$ and NDCG@$K$ focus more on ranking quality. When there is only one ground-truth item (as in our setting), MRR@$K$ and NDCG@$K$ only differ in the sensitivity concerning the ranking. Since the ranking score discounts logarithmically in NDCG, it decreases much slower than in MRR. In this case, items hit at the back of the recommendation list can still contribute considerably to the performance.

\subsubsection{Hyper-parameters}
For fair comparison, following {the} previous works~\cite{li2017neural,liu2018stamp,wu2019session}, we set the embedding size $d = 100$ for both the unsupervised pre-training process and the supervised learning process. Model parameters are initialized using a Gaussian distribution with a mean of 0 and a standard deviation of 0.1. We use the mini-batch Adam optimizer to optimize model parameters {by setting the initial learning rate as $1e-3$}. The default values of hyper-parameters for \emph{node2vec} is shown in Table~\ref{tab:node2vec}. We only modify the return parameter $p$ and the in-out parameter $q$ in the following experiments. Other hyper-parameters are {tuned on the validation set}.

\subsection{Comparison Results}
\label{subsec:comparison}

To demonstrate the overall performance of the proposed G$^3$SR method, we compare it with the eleven baselines described in Section~\ref{subsec:setup} on three datasets. The {experimental results are} presented in Table~\ref{tab:comp} and we bold the top-2 results\footnote{{Experimental results for different choices of {$K$} are reported in Appendix~\ref{app:topk}.}}. We also show the improvement w.r.t. the GRU4REC~\cite{hidasi2016session} model for each other method. As we can see from the table, G$^3$SR achieves {very promising results on all three datasets. Remark} that the test set for Yoochoose 1/64 and Yoochoose 1/4 are the same. In general, a model trained on a larger amount of data (1/4) performs better than a model trained on {fewer} data (1/64), but here we observe an adverse effect. On the one hand, we notice that using more training data does not necessarily improve performance since it may also introduce noisy information and can not reflect user's recent interests precisely. On the other hand, G$^3$SR achieves 72.61 in P@20 with merely 1/64 training data, even higher than the performance of {all other models trained on 1/4 training data, including the two contemporary approaches FGNN and GCE-GNN}. This significantly reduces the training time and verifies the effectiveness of G$^3$SR for \emph{using fewer data to achieve better results}.

Over the traditional recommendation methods, Item-KNN achieves the best results, even better than the FPMC method which takes item transition into consideration. This reveals that the independence assumption widely used in MC-based methods is too strong. As one of the most popular recommendation methods, BPR works poorly on all three datasets, especially on Yoochoose 1/4, even worse than the simple S-POP method. Overall, the experimental results show that traditional methods fail to fit the session-based recommendation problem since the identities of users are missing and contextual information are unconsidered.

\begin{table}[!t]
\centering
    \caption{Default settings for ablation studies and sensitivity analyses.}
    \vskip -0.1in
    \label{tab:default}
    \footnotesize
    \centering
    \begin{tabular}{lc}
        \toprule
        Factor & Setting\\
        \midrule
        readout strategy&exponential decaying (Eq.~(\ref{eq:expReadout}))\\
        pre-train dataset&Yoochoose 1/64 or Diginetica\\
        temperature parameter $\tau$ & 1.0\\
        \# of layers $T$ & 1\\
        return parameter $p$ & 0.25\\
        in-out parameter $q$ & 4\\
        \bottomrule
    \end{tabular}
\end{table}

Deep-learning-based methods consistently outperform traditional methods, owing to their extraordinary representation learning ability for sequential data. GRU4REC and NARM use RNN to learn behavioral patterns in session data and achieve decent performance. NARM also uses the attention mechanism to learn user's general interest which significantly improves the performance. STAMP abandons the RNN model and adopts simple MLP architecture enhanced by a well designed short-term priority attention mechanism. This frees the model from calculating the session representation sequentially and therefore makes better use of the parallel computing ability of GPUs. Overall, this part of experiments demonstrates that it is highly important to model the complete session sequence instead of merely considering the last click. However, both RNN and MLP fail to capture the complex transitions between items in session data. This is why they work worse than graph-aware methods.

{For the graph-aware methods, we can see that they outperform other baselines by a large margin. In particular, G$^3$SR outperforms SR-GNN on all three metrics for the Diginetica dataset. For the Yoochoose dataset, it seems that SR-GNN performs slightly better on MRR@20 while G$^3$SR has better or competitive performance on P@20 and NDCG@20. Compared {with} the two contemporary approaches that also utilize global information, namely FGNN and GCE-GNN, G$^3$SR achieves competitive results or even performs better, {especially when comparing the results on Yoochoose 1/64 with those on Yoochoose 1/4}. These results further affirms the effectiveness of the proposed method.}

{Remark that recommendation is usually decomposed into two phases in the industrial community~\cite{covington2016deep}, i.e., recall and ranking. The recall phase aims to accurately shortlist candidate items from the full item set with relatively fewer features. In contrast, the ranking phase introduces fine-grained features and refines the ranking quality based on the results provided by the recall phase. Higher precision is of significance for the recall phase since it implies the model can accurately recall candidate items. The proposed G$^3$SR method deals with the full item set and achieves a high precision with only the ID feature, which makes it a good recall model. What's more, the pre-trained embeddings can be useful for many downstream tasks~\cite{wang18billion,grbovic18real}.}

{To validate this experimental result, we perform detailed ablation studies and sensitivity analyses in Section~\ref{subsec:ablation} and Section~\ref{subsec:sensitivity} respectively. Last but not least, {in Appendix~\ref{app:model}, we study the performance of our method when used with other model structures, e.g., GC-SAN~\cite{xu2019graph} and GCE-GNN~\cite{wang2020}.

{For clarity, we use the default settings shown in Table~\ref{tab:default}{, which use the same data volume and model structure as in~\cite{wu2019session}} to conduct experiments in the following sections and only modify one of the factors in each experiment.}

\begin{table}
\centering
\caption{The performance of G$^3$SR with different readout strategies.}
\vskip -0.1in
\label{tab:readout}
\footnotesize
\resizebox{0.5\textwidth}{!}{
\begin{tabular}{@{}lcccccc@{}}
\toprule
\multirow{2}{*}{Method} & \multicolumn{3}{c}{Yoochoose 1/64} & \multicolumn{3}{c}{Diginetica}\\
&P@20    &MRR@20  &NDCG@20  &P@20    &MRR@20   &NDCG@20\\
\midrule
{SR-GNN}          & {70.57}    & {30.94}   & 39.20  & {50.73}    & {17.59} & 24.68\\
\midrule
G$^3$SR-last    & 71.03    & 30.53   & 39.77  & 52.86    & 18.45 & 26.08\\
G$^3$SR-mean    & 67.85    & 27.48   & 36.62  & 51.67    & 17.96 & 25.40\\
G$^3$SR-sum     & 68.55    & 28.45   & 37.57  & 51.99    & 17.71 & 25.29\\
G$^3$SR-attn    & 71.44    & \textbf{30.83}   & \textbf{40.10}  & 53.18    & \textbf{18.71} & \textbf{26.31}\\
G$^3$SR         & \textbf{71.58}    & 30.33   & 39.71  & \textbf{53.62}    & 18.52 & 26.30\\
\bottomrule
\end{tabular}}
\end{table}

\begin{table}
\centering
\caption{The performance of G$^3$SR and G$^3$SR-attn with different size of pre-training data.}
\vskip -0.1in
\label{tab:pre-training}
\footnotesize
\resizebox{0.5\textwidth}{!}{
\begin{tabular}{@{}lcccccc@{}}
\toprule
\multirow{2}{*}{Proportion} & \multicolumn{3}{c}{G$^3$SR} & \multicolumn{3}{c}{G$^3$SR-attn}\\
&P@20    & MRR@20  & NDCG@20  & P@20    & MRR@20  & NDCG@20\\
\midrule
1/1    & 72.51    & 30.67   & 40.18  & 72.20    & \textbf{31.51} & 40.76\\
1/2    & 72.31    & 30.70   & 40.21  & \textbf{72.34}    & 31.50 & \textbf{40.80}\\
1/4    & \textbf{72.55}    & 30.73   & \textbf{40.27}  & 72.25   & 31.26 & 40.50\\
1/8    & 72.43    & \textbf{30.76}   & 40.26  & 72.16    & 31.06 & 40.42\\
1/16   & 72.24    & 30.58   & 40.10  & 72.17    & 30.93  & 40.34\\
1/32   & 72.02    & 30.60   & 40.05  & 71.65    & 30.91  & 40.19\\
1/64   & 71.58    & 30.33   & 39.71  & 71.44    & 30.83  & 40.10\\
\bottomrule
\end{tabular}}
\end{table}

\subsection{Ablation Studies}
\label{subsec:ablation}

\subsubsection{Performance of different readout mechanism{s}}
\label{subsubsec:perf_readout}
Although we only describe the exponential decaying readout strategy in Section~\ref{subsubsec:readout}, G$^3$SR can be modified to adopt various readout {strategies} such as mean, summation and the widely used attention mechanism. In this section, we compare the proposed method with its variants that use different readout functions:
\begin{itemize}
\item G$^3$SR-last readouts the last click only.
\item G$^3$SR-mean readouts the average of item representations.
\item G$^3$SR-sum readouts the summation of item representations and passes it through a linear layer.
\item G$^3$SR-attn uses the short-term priority attention mechanism~\cite{liu2018stamp,wu2019session} to weight item representations.
\end{itemize}
The experimental result is shown in Table~\ref{tab:readout}. First, we can see that all the four variants achieve competitive results, confirming the validity of the unsupervised pre-training process. Among these variants, only G$^3$SR-sum and G$^3$SR-attn require additional model parameters. According to the results of G$^3$SR-mean and G$^3$SR-sum, we can see that permutation invariant readout strategies have the worst results because it regards all nodes (items) {as having} equal importance, which is easily affected by noisy data and violates the nature of the session-based recommendation problem ----- user's interests drift along with time. G$^3$SR-last performs fairly well since the message passing process has delivered information of other items to the last clicked item. Overall, we can see that G$^3$SR-attn and G$^3$SR perform the best with the former wins {on the ranking metrics and the latter wins on precision}. This further affirms the experimental results we observed in Table~\ref{tab:comp}. The reason why SR-GNN performs better in MRR@20 is {that it uses the attention mechanism}, which can be easily equipped to the proposed G$^3$SR method as well. {Note} that the proposed exponential decaying strategy requires neither additional model parameters nor matrix multiplications but still beats the attention-based strategy in P@20 which implies that it captures user's potential interests more precisely.

\begin{table}
\centering
\caption{The effect of the pre-training process and the task-specific bias.}
\label{tab:pre-training2}
\footnotesize
\resizebox{0.5\textwidth}{!}{
\begin{tabular}{@{}lcccccc@{}}
\toprule
\multirow{2}{*}{Method} & \multicolumn{3}{c}{Yoochoose 1/64} & \multicolumn{3}{c}{Diginetica}\\
&P@20    & MRR@20   & NDCG@20   & P@20    & MRR@20    & NDCG@20\\
\midrule
G$^3$SR    & \textbf{71.58}    & \textbf{30.33}   & \textbf{39.71}  & \textbf{53.62}    & \textbf{18.52} & \textbf{26.30}\\
G$^3$SR w/o pre-train    & 70.14    & 30.32   & 39.41  & 50.59    & 17.60 & 24.90\\
G$^3$SR w/o bias  & 66.95    & 27.94   & 36.78  & 49.16    & 17.41 & 24.40\\
\bottomrule
\end{tabular}}
\end{table}

\begin{figure}[!t]
    \centering
    \includegraphics[width=1.0\linewidth]{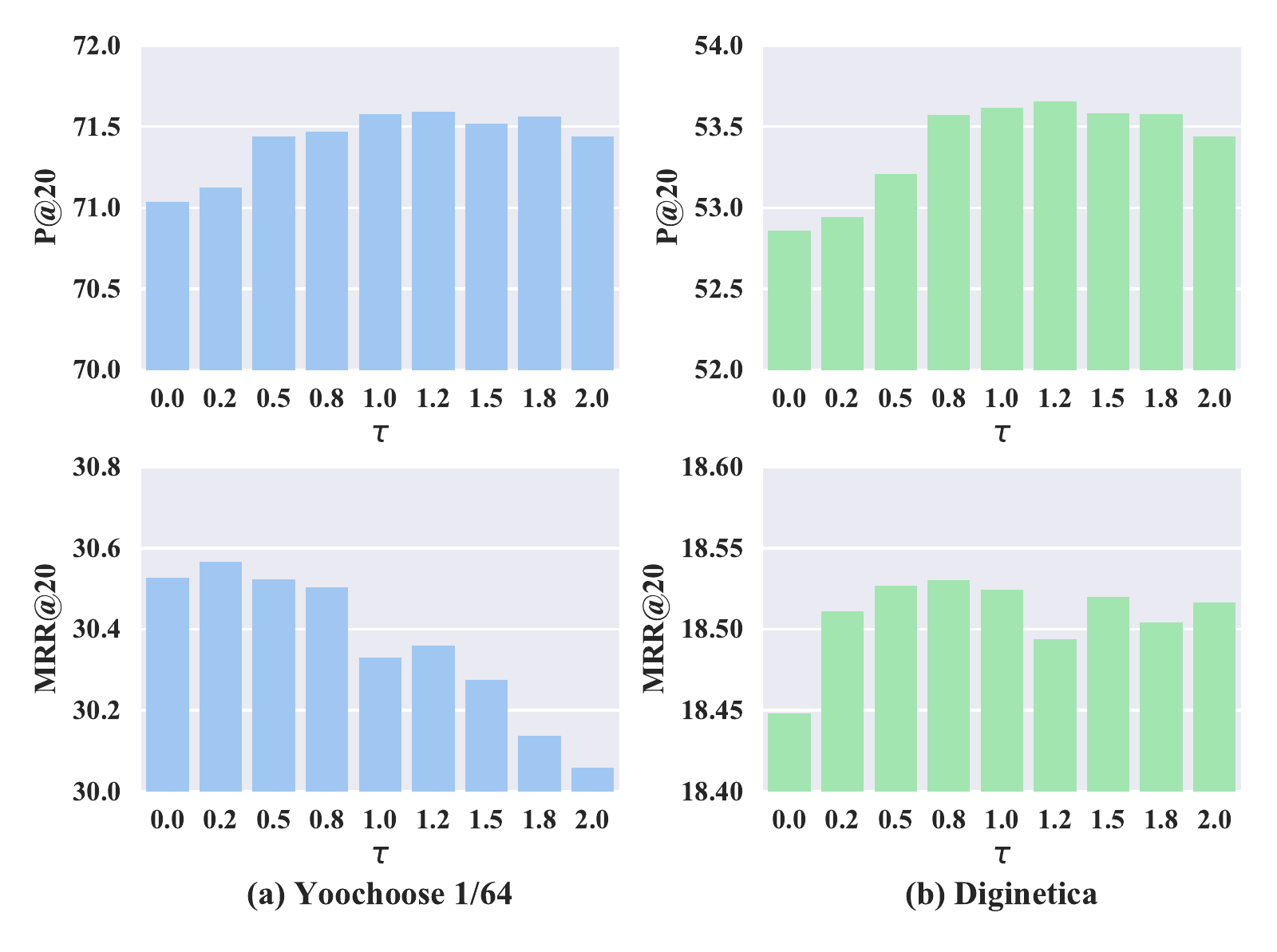}
    \caption{The effect of the temperature parameter $\tau$.}
    \label{fig:temparature}
\end{figure}

\begin{figure}[!t]
    \centering
    \includegraphics[width=1.0\linewidth]{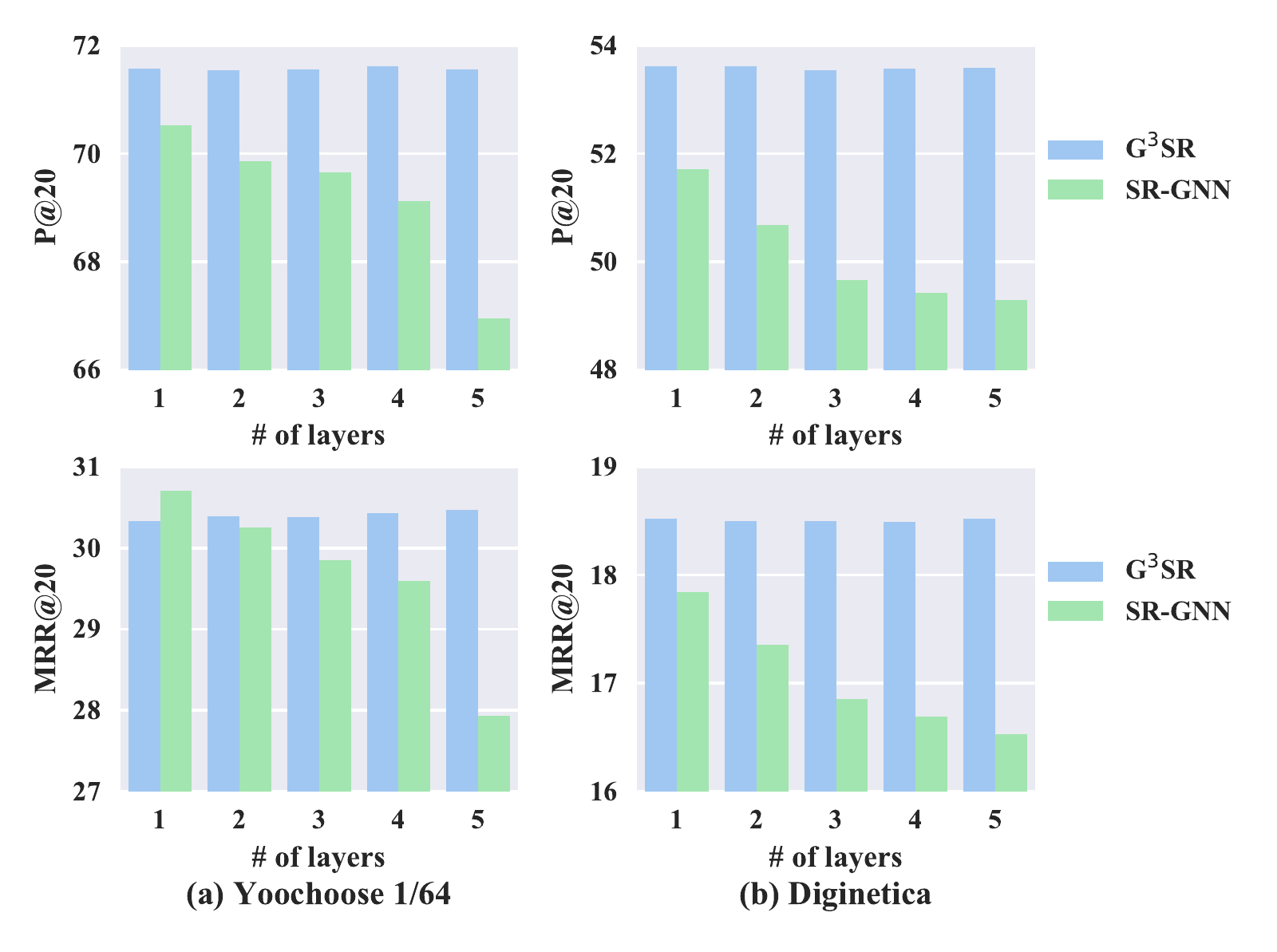}
    \caption{The effect of the number of layers $T$.}
    \label{fig:layers}
\end{figure}

\begin{figure}[!t]
    \centering
    \includegraphics[width=0.95\linewidth]{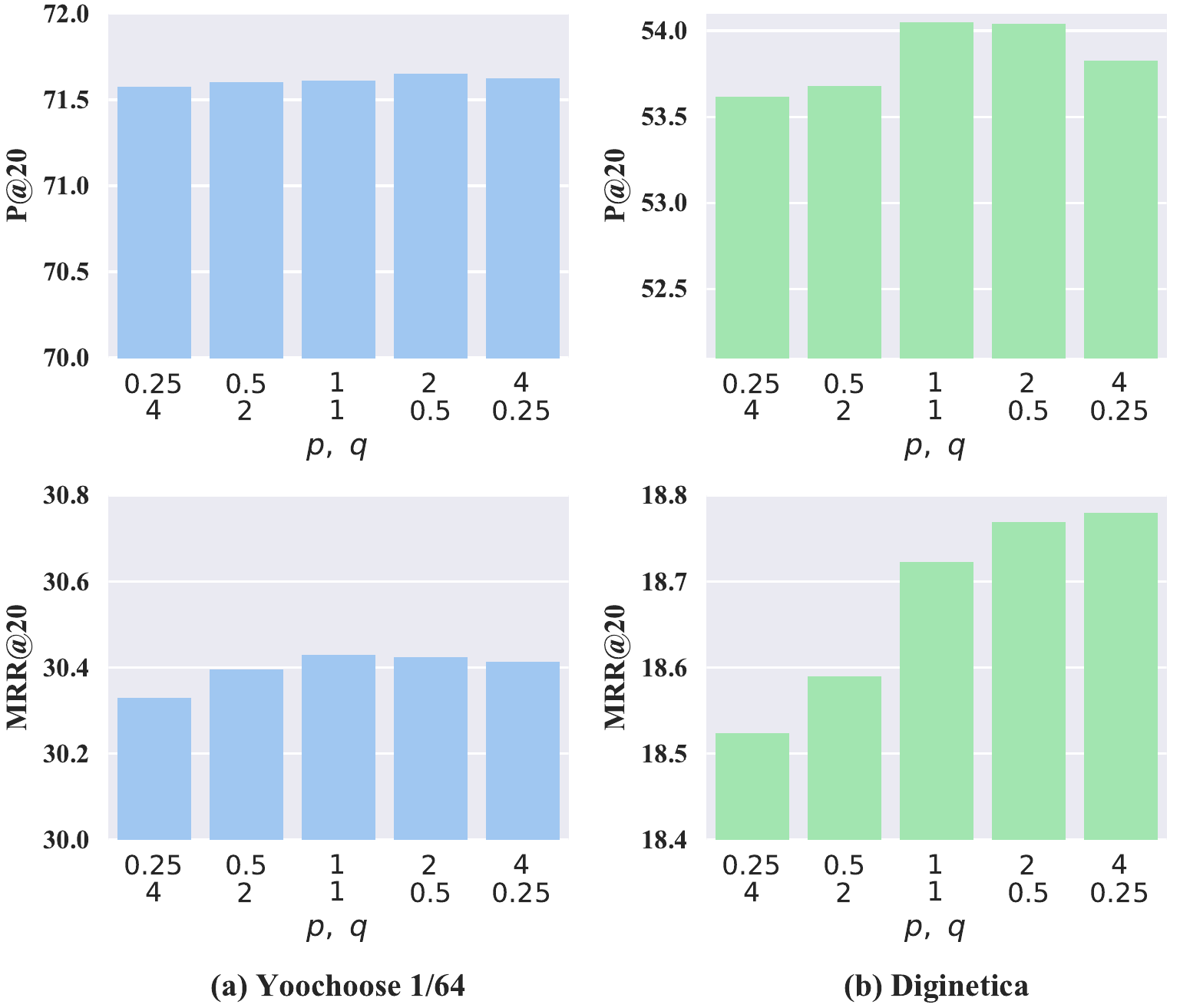}
    \caption{The effect of the return parameter $p$ and the in-out parameter $q$.}
    \label{fig:pq}
\end{figure}

\subsubsection{Pre-training on larger datasets}
In a real-world situation, companies usually have massive available data though they tend to use the most recent part in order to keep up with users' interests changing in time. The above experiments use the same size of data in both the unsupervised pre-training process and the supervised learning process. However, one of the biggest advantages of the proposed method is that the pre-training process can make use of more available data to learn invariant relational information while the model for downstream task can still be trained on recent data to adapt to the changes. In this section, we explore the possibility of using larger datasets for pre-training. Specifically, we use different proportion of the Yoochoose dataset to pre-train item embeddings and then train the recommendation model with the most recent 1/64 of data. The experimental results are shown in Table~\ref{tab:pre-training}.

As we can see from the table, the effectiveness of pre-training on larger dataset is obvious. We perform experiments for both the G$^3$SR and G$^3$SR-attn methods and the improvements are consistent. The use of pre-training further pushes P@20 to a new level for the Yoochoose dataset, i.e., higher than 72\%. It's clear that G$^3$SR performs better than G$^3$SR-attn {on precision and worse on ranking}, again confirming the observations mentioned above. The best performance for G$^3$SR is achieved with 1/4 of data while the counterpart for G$^3$SR-attn is 1/2. This is reasonable since G$^3$SR-attn has a larger model capacity and can learn importance weights adaptively. Overall, the experimental results indicate that pre-training on larger datasets is useful but it is unnecessary to use all available data.

\subsubsection{Pre-training and task-specific bias learning}
As we describe in Section~\ref{subsubsec:initialization}, the node representation in a session graph is constructed by adding a task-specific bias $\beta(v)$ to the corresponding item embedding $\phi(v)$ learned in the unsupervised pre-training process. To validate whether the improvements come from pre-training, task-specific bias or both, we conduct further experiments that remove these features {and the results are shown in Table~\ref{tab:pre-training2}. Specifically, G$^3$SR w/o pre-train uses the same number of parameters but removes the pre-training process. G$^3$SR w/o bias removes the task-specific bias and fine-tunes the item representations obtained in the unsupervised pre-training process.}

{From Table~\ref{tab:pre-training2}, w}e can learn that removing any one of these two features significantly downgrades the performance. If we remove the pre-training process, the model fails to perceive the global graph structure and therefore miss some key information that is useful for reasoning on session graphs. If we remove the task-specific bias, the performance drops dramatically because the goals in the two learning processes are essentially different, i.e., the unsupervised pre-training aims to predict contextual items and the {supervised} learning process aims to predict the next item. {These} experimental results show that both the pre-training process and the task-specific bias are of significance in the proposed G$^3$SR method.

\subsection{Sensitivity Analyses of Hyper-parameters}
\label{subsec:sensitivity}

\subsubsection{Temparature parameter $\tau$}
In Eq.~(\ref{eq:expReadout}), we use the temperature parameter $\tau$ to control the degree of attenuation. With a small value of $\tau$, the weights are mostly distributed to the last few clicks. We are interested in how sensitive the proposed G$^3$SR method is to the temperature parameter $\tau$.

\begin{figure*}[!t]
    \centering
    \includegraphics[width=1.0\linewidth]{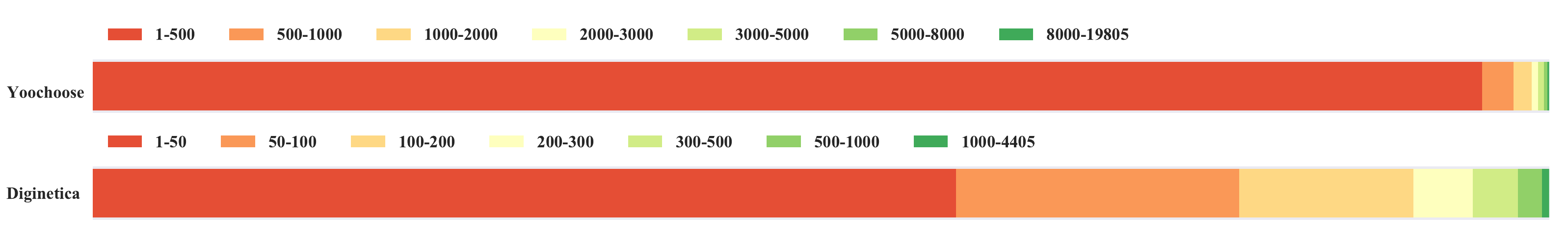}
    \caption{The distributions of item popularity for the Yoochoose dataset and the Diginetica dataset.}
    \label{fig:popularity}
\end{figure*}

The experimental results are shown in \figurename~\ref{fig:temparature}. Note that, if we set $\tau=0$, the weight of the last clicked item goes to positive infinity. In this case, the exponential decaying strategy downgrades to G$^3$SR-last which only readouts the last click from the session graph. On the contrary, if $\tau \rightarrow +\infty$, G$^3$SR downgrades to G$^3$SR-mean which reads out the average representation of all items in the session graph. As shown in the table, the best P@20 is achieved near $\tau=1.2$ on both datasets while the observation in MRR@20 shows a slightly different trend. We see that the performance drops quickly along with the increase of $\tau$ on the Yoochoose 1/64 dataset while the Diginetica dataset seems to be less sensitive to the change of $\tau$. According to Eq.~(\ref{eq:expReadout}), the weight for each position depends on both $\tau$ and the session length. Therefore, the difference in the distribution of session length accounts for this observation. Specifically, the Yoochoose 1/64 dataset has longer max length as well as average length than the Diginetica dataset. To achieve better ranking performance, the Yoochoose 1/64 dataset requires a smaller $\tau$ so that larger probabilistic density can accumulate on the last few clicks. Overall, a larger $\tau$ leads to more accurate recommendation since it allows the model to {capture} more useful contextual information, but an overlarge $\tau$ is inappropriate as it treats every item equally. Besides, increasing $\tau$ may also lead to worse ranking performance.

\subsubsection{Number of layers $T$}
Except for the temperature parameter $\tau$, we are also interested in the number of graph network layers $T$~(the total time steps in the message passing process). Larger value of $T$~(more time steps) implies that a node can learn from its higher-order neighbors. Therefore, it is of significance for large and complex graph structured data.

Figure~\ref{fig:layers} shows the experimental results for G$^3$SR and SR-GNN with different number of network layers. It is clear that, with the increase of the number of layers, G$^3$SR performs stably with small improvements while the performance of SR-GNN drops dramatically in both metrics. Unlike general neural networks such as MLP, the total number of model parameters of a graph network does not increase with more layers since all time steps share the same parameters. As a result, {using} more layers does not necessarily incur the notorious over-fitting problem. Instead, deep graph networks may easily fall into the trap of the over-smoothing problem~\cite{li2018deeper, li2019deepgcns} where node features tend to converge to the same value. The experimental results verify that the proposed G$^3$SR method can {alleviate} over-smoothing on some level {since the fixed representation $\phi(v)$ learned in the pre-training process serves as an unchangeable part of the node representation during the message passing process. Note that, although this method is not exactly the same as  the popular techniques used in GNNs to avoid over-smoothing, e.g., residual connections and multi-hop methods~\cite{kipf2016semi, xu2018representation, li2019deepgcns, abu-el-haija2019n}. They follow the same principle, i.e., maintaining the distinguishing ability of the node representations.}

\subsubsection{Return parameter $p$ and in-out parameter $q$}
The return parameter $p$ and the in-out parameter $q$ are introduced in the unsupervised pre-training process to control the learning of item embeddings. More specifically, as discussed in~\cite{grover2016node2vec}, $p$ and $q$ control how fast the walk explores and leaves the neighborhood of a source node. They allow the random walk procedure to approximately interpolate between Breadth-First Sampling (BFS) and Depth-First Sampling (DFS) and therefore affect the contextual nodes in the objective function, i.e., Eq.~(\ref{finalObjective}). Note that, \emph{node2vec} degenerates to \emph{DeepWalk}~\cite{perozzi2014deepwalk} when $p=q=1$, as all the neighborhood nodes are sampled with equal probability.

The experimental results of different combinations of the return parameter $p$ and the in-out parameter $q$ are shown in \figurename~\ref{fig:pq}. We can see that the effects on different datasets are quite different. Changing $p$ and $q$ doesn't seem to affect the performance significantly on the Yoochoose dataset. Both P@20 and MRR@20 remain level with the increase of $p$ and the decrease of $q$. In contrast, an appropriate choice of the hyper-parameter $p$ and $q$ can improve the performance on Diginetica. The best P@20 is achieved when $p=2, q=0.5$ and we can observe an upward trend in MRR@20 with the increase of $p$ and the decrease of $q$. Such differences can be attributed to the differences in the distribution of item popularity as we will discuss in Section~\ref{subsec:case}. The distribution of Yoochoose has a heavy tail, i.e., the vast majority of items are rarely clicked, which greatly restricts the performance of explorations (larger $p$ and smaller $q$). The distribution of Diginetica, by contrast, is much smoother and therefore can easily benefit from extra explorations.

\subsection{Case Studies}
\label{subsec:case}
\subsubsection{Item popularity}
As stated in Section~\ref{subsubsec:learn_embedding}, we use random walks instead of session sequences to learn item embeddings in the unsupervised pre-training process and we argue that cold items can benefit from this process. In this section, we conduct experiments to further study the effect of the proposed method.
\begin{table}[!t]
\centering
\caption{The number of cold items and popular items in different datasets.}
\label{tab:partitions}
\begin{tabular}{@{}lcccc@{}}
\toprule
Partition & Yoochoose & Diginetica\\
\midrule
\textbf{cold items} & 13,336 & 16,961\\
\textbf{popular items} & 13,078 & 17,280\\
\bottomrule
\end{tabular}
\end{table}

\begin{table}[!t]
\centering
\caption{The performance on cold items and popular items.}
\label{tab:cold}
\footnotesize
\resizebox{0.5\textwidth}{!}{
\begin{tabular}{lcccccc}
\toprule
\multirow{2}{*}{Method} & \multicolumn{3}{c}{Yoochoose} & \multicolumn{3}{c}{Diginetica}\\
&P@20    & MRR@20   & NDCG@20   & P@20    & MRR@20   & NDCG@20\\
\midrule
\multicolumn{7}{c}{cold items}\\
\midrule
G$^3$SR      &\textbf{59.60} &\textbf{25.56} & \textbf{33.28} & 34.28 &12.30 & 17.08 \\
G$^3$SR-attn &59.11 &25.47 & 33.10 &\textbf{34.36} &\textbf{12.43} & \textbf{17.21} \\
SR-GNN       &53.81 &23.21 & 30.15 &32.42 &9.54 & 14.51 \\
\midrule
\multicolumn{7}{c}{popular items}\\
\midrule
G$^3$SR      &\textbf{86.05} &40.04 & 50.79 &\textbf{69.76} &\textbf{25.22} & \textbf{35.28} \\
G$^3$SR-attn &85.95 &41.09 & 51.55 &68.67 &25.19 & 35.01 \\
SR-GNN       &85.79 &\textbf{41.54} & \textbf{51.85} &66.66 &25.21 & 34.51 \\
\bottomrule
\end{tabular}}
\end{table}
We first {analyse} the distribution of item popularity (the number of occurrence in the training set) for the Yoochoose dataset and the Diginetica dataset. As shown in \figurename~\ref{fig:popularity}, the long tail phenomenon is observed on both datasets as cold items dominate the session data. Since the range of item popularity varies from one dataset to another, we use different scales for the two datasets to depict the figures. For the Yoochoose dataset, the item popularity ranges from 1 to 19,805 and over 95\% of items occur less than 500 times. In this case, we refer to items with popularity lower than or equal to 500 as cold items and items with popularity higher than 5,000 as popular items. The thresholds for the Diginetica dataset is set to 50 and 300 respectively, smaller than its counterparts for the Yoochoose dataset since its distribution of item popularity is much narrower. About 60\% of items in the Diginetica dataset occur less than 50 times and 20\% of items occur between 50 times and 100 times. Overall, we can see that both datasets obey the long tail distribution and the Diginetica dataset has a smoother distribution than the Yoochoose dataset.
\begin{table}[!t]
\centering
\caption{The number of long sessions and short sessions in different datasets.}
\label{tab:partitions2}
\begin{tabular}{@{}lcccc@{}}
\toprule
Partition & Yoochoose & Diginetica\\
\midrule
\textbf{long sessions} & 16,713 & 14,363\\
\textbf{short sessions} & 39,185 & 46,495\\
\bottomrule
\end{tabular}
\end{table}
To validate the effectiveness of the proposed method, we split the test set according to the item popularity of label items. For example, the threshold of cold items for Yoochoose is 500. If the label item $v_{|\mathbf{s}^{(i)}|+1}^{(i)}$ for session $\mathbf{s}^{(i)}$ occurs less than 500 times in the training set then $\mathbf{s}^{(i)}$ is partitioned into \textbf{cold}. The number of sessions in different partitions is shown in Table~\ref{tab:partitions}.

{Table}~\ref{tab:cold} shows the comparison results of G$^3$SR, G$^3$SR-attn and SR-GNN on the two datasets in terms of cold items and popular items. Generally, the proposed methods (both G$^3$SR and G$^3$SR-attn) achieve better results than SR-GNN for cold items and popular items. This again confirms the effectiveness of the proposed methods. Furthermore, the proposed method outperforms SR-GNN in both metrics when recommending cold items, especially on the Yoochoose dataset where the long tail phenomenon is much more severe. The underlying reason is that the random walks used in the unsupervised pre-training process augment session data for cold items, which is useful for learning good semantic representations. Although the proposed methods work well in most cases, we should also notice the exception on the Yoochoose dataset when recommending popular items. {SR-GNN is competitive in P@20 and is good at ranking popular items}. This reveals that for datasets that are dominated by cold items, improving the performance on recommending cold items may sacrifice the {ranking performance} on popular items, which also explains the experimental results shown in Table~\ref{tab:comp}. In practice, people need to strike the balance carefully to improve the overall performance of the system.
\begin{table}[!t]
\centering
\caption{The performance on long sessions and short sessions.}
\label{tab:length}
\footnotesize
\resizebox{0.5\textwidth}{!}{
\begin{tabular}{lcccccc}
\toprule
\multirow{2}{*}{Method} & \multicolumn{3}{c}{Yoochoose} & \multicolumn{3}{c}{Diginetica}\\
&P@20    & MRR@20    & NDCG@20    & P@20    & MRR@20    & NDCG@20\\
\midrule
\multicolumn{7}{c}{long sessions}\\
\midrule
G$^3$SR      &\textbf{68.08} &25.46 & 35.12 &\textbf{51.35} &\textbf{16.38} & \textbf{24.09} \\
G$^3$SR-attn &67.90 &\textbf{25.98} & \textbf{35.49} &50.70 &16.29 & 23.87 \\
SR-GNN       &66.65 &25.51 & 34.81 &49.02 &15.55 & 22.91 \\
\midrule
\multicolumn{7}{c}{short sessions}\\
\midrule
G$^3$SR      &\textbf{73.06} &32.45 & 41.77 &\textbf{54.87} &\textbf{19.51} & \textbf{27.36} \\
G$^3$SR-attn &72.96 &\textbf{32.88} & \textbf{42.06} &53.94 &19.39 & 27.07 \\
SR-GNN       &72.19 &32.77 & 41.78 &52.55 &18.56 & 26.06 \\
\bottomrule
\end{tabular}}
\end{table}
\subsubsection{Session Length}
{As shown in {Section}~\ref{subsubsec:perf_readout}, the simple exponential decaying readout strategy performs better than the attention mechanism. Seeing as G$^3$SR-last works fairly well, the basic assumption used in~\cite{liu2018stamp,wu2019session} is empirically true, i.e., the last clicked/viewed item is usually more relevant to the next item. However, these papers also claim that using the attention mechanism can significantly improve the recommendation performance for long sessions, which is not necessarily true according to our experiments. To get a better understanding of the performance improvements, we conduct experiments to further analyse the effect of session length in this section.}

{Each test set is divided into two subsets according to the length of each session. Sessions longer than 5 are regarded as long sessions and the rest are short sessions. The statistics for the partition is given in {Table}~\ref{tab:partitions2}. We can see that short sessions dominate both of the datasets, accounting for up to 70.1\% and 76.4\% respectively. This is also true in real-world scenarios~\cite{wang19serendipitous} since most users are unlikely to spend too much time on the website and tend to browse only a few webpages.}

{{Table}~\ref{tab:length} shows the experimental results. First, we can see that the recommendation performance for short sessions is consistently better than long sessions, which is true for all the considered methods. Second, G$^3$SR performs better than G$^3$SR-attn and SR-GNN, and the magnitude of improvements is similar in both cases, i.e., the exponential decaying strategy is effective for both long sessions and short sessions. Intuitively, an attention-based method should perform better on long sessions since it's more flexible and has a higher model capacity, but the results show the opposite. Although the superiority on the Yoochoose dataset is less obvious, it still proves that a simple non-parametric readout strategy that fits the nature of the problem has the potential to be a better choice, especially for its simplicity. Note that we are not claiming that attention-based readout methods are useless but aiming to call for reconsidering the necessity and how to use them to better fit the nature of the problem. According to the results, we can draw a safe conclusion that, after the {pre-train}ing process and the message passing process, the information accumulated in the representations of the last few items is already enough for predicting the next item, which makes attention-based readout methods less attractive.}

\section{Conclusion}
\label{sec:conclusion}
In this work, we propose a simple but strong baseline for the session-based recommendation problem called Global Graph Guided Session-based Recommendation (G$^3$SR). It decomposes the problem into two processes, i.e., the unsupervised pre-training process and the supervised learning process. The remarkable and consistent improvements show some interesting insights. First, learning {global information through pre-training is of significance for improving the performance of session-based recommendation}, especially for cold items. Second, pre-training on larger dataset{s is useful and has great} potential. Although we use a simple node2vec algorithm to pre-train item embeddings {in this paper}, better pre-training techniques are to be explored. Last but not least, as we discussed in Section~\ref{sec:experiments}, the proposed exponential decaying strategy is simple but highly competitive with the sophisticated attention mechanism. {As a result, it's promising to design better representation learning methods and/or readout functions for session-based recommendation in future works.}

% or


%\appendices
%% you can choose not to have a title for an appendix
%% if you want by leaving the argument blank
%\section{}
\begin{table}[!t]
\centering
\caption{Performance comparison between SR-GNN, G$^3$SR-attn and G$^3$SR over three benchmark datasets}
\label{tab:topk}
\begin{threeparttable}
\resizebox{0.5\textwidth}{!}{
\begin{tabular}{@{}l@{}cccccc@{}}
\toprule
\multirow{2}{*}{Method} & \multicolumn{2}{c}{Yoochoose 1/64} & \multicolumn{2}{c}{Yoochoose 1/4} & \multicolumn{2}{c}{Diginetica} \\
&P@{$K$}    & MRR@{$K$}    & P@{$K$}    & MRR@{$K$}    & P@{$K$}    & MRR@{$K$}\\
\midrule
\multicolumn{7}{c}{{$K=1$}}\\
\midrule
SR-GNN      & \ \textbf{17.34}    & \textbf{17.34}    & \textbf{18.89}    & \textbf{18.89}    & 9.00    & 9.00\\
G$^3$SR-attn     & 17.00    & 17.00    & 18.22    & 18.22    & \textbf{9.41}    & \textbf{9.41}\\
G$^3$SR     & \ 16.54    & 16.54    & 17.00    & 17.00    & 9.32    & 9.32\\
\midrule
\multicolumn{7}{c}{{$K=5$}}\\
\midrule
SR-GNN      & \ 47.11    & 28.13    & 48.09    & \textbf{29.44}    & 27.34    & 15.43\\
G$^3$SR-attn     & 48.05    & \textbf{28.34}    & \textbf{48.18}    & 29.04    & 28.89    & 16.26\\
G$^3$SR     & \ \textbf{48.38}    & 28.16    & 47.76    & 28.15    & \textbf{29.10}    & \textbf{16.30}\\
\midrule
\multicolumn{7}{c}{{$K=10$}}\\
\midrule
SR-GNN      & \ 59.95    & 29.85    & 60.90    & \textbf{31.16}    & 38.97    & 16.96\\
G$^3$SR-attn     & 60.99    & \textbf{30.08}    & \textbf{61.02}    & 30.76    & 40.37    & 17.77\\
G$^3$SR     & \ \textbf{61.80}    & 29.96    & 60.88    & 29.92    & \textbf{40.85}    & \textbf{17.86}\\
\midrule
\multicolumn{7}{c}{{$K=20$}}\\
\midrule
SR-GNN      & \ 70.57    & \textbf{30.94}    & 71.36    & \textbf{31.89}    & 50.73    & 17.59\\
G$^3$SR-attn     & 71.44    & 30.83    & 71.14    & 31.88    & 53.18    & 18.71\\
G$^3$SR     & \ \textbf{72.61}    & 30.74    & \textbf{71.62}    & 30.68    & \textbf{54.05}    & \textbf{18.72}\\
\bottomrule
\end{tabular}}
\begin{tablenotes}
\item[*] Note that MRR@{$K$} is exactly the same with P@{$K$} when {$K=1$}, i.e., the ranking metric degenerates and only cares about whether the predicted item is correct or not in this case.
\end{tablenotes}
\end{threeparttable}
\end{table}
\appendices
\section{A closer look at the top-K results}
\label{app:topk}
{To provide a better understanding of the proposed methods, we compare SR-GNN, G$^3$SR-attn and G$^3$SR on the three datasets w.r.t. different choices of K and the experimental results are presented in {Table}~\ref{tab:topk}. Overall, the results suggest a similar trend as observed in {Table}~\ref{tab:comp}. The proposed methods consistently perform better than SR-GNN except for the top-1 cases. The experiments in Section~\ref{subsubsec:perf_readout} have revealed that attention-based readout function {tends} to have better ranking performance so its strength is gradually shown when we decrease {$K$}, which explains why SR-GNN and G$^3$SR-attn perform better for smaller {$K$}. In practice, we need to strike the balance between precision and ranking performance according to the specific {requirements}. For example, a simple model {having} high precision is in need {when recalling} candidate items from all items, but a model with better ranking performance is preferable {to generate the final recommendation list}. Seeing as G$^3$SR's high precision and simplicity, it has the potential to be a good recall model.}
\vskip -0.2in
\begin{table}[!t]
\centering
\caption{Performance comparison {with different model structures}}
\label{tab:GCE}
\begin{threeparttable}
\resizebox{0.5\textwidth}{!}{
\begin{tabular}{lcccc}
\toprule
\multirow{2}{*}{Method} & \multicolumn{2}{c}{Yoochoose 1/64} & \multicolumn{2}{c}{Diginetica} \\
& P@20    & MRR@20    & P@20    & MRR@20\\
\midrule
SR-GNN             & 70.36    & 30.73    & 50.73    & 17.59\\
G$^3$SR            & 72.61    & \textbf{30.83}    & 54.05    & 18.72\\
{GC-SAN}             & 68.49    & 29.54    & 42.75    & 13.86\\
{G$^3$SR+GC-SAN}     & 71.90    & 30.82    & 52.30    & 18.05\\
GCE-GNN            & 71.75    & 30.43    & 54.18    & \textbf{19.01}\\
G$^3$SR+GCE-GNN    & \textbf{72.83}    & 30.78    & \textbf{54.29}    & 18.94\\
\bottomrule
\end{tabular}}
\end{threeparttable}
\end{table}
\section{{Comparison with other model structures}}
\label{app:model}
{In this section, we study the performance of our method when used with other model structures. Specifically, GC-SAN~\cite{xu2019graph} and GCE-GNN~\cite{wang2020} are considered here. The former introduces the self-attention mechanism to better exploit intra-session information. The latter is currently the best performing algorithm, which is written in the same timeframe with our work and also takes global graph into consideration. These methods are orthogonal to our research since they focus on designing better model structures while our method pre-trains item embeddings on global graph and strives to use it to simplify the downstream task. Therefore, G$^3$SR can be flexibly combined with both GC-SAN and GCE-GNN.}

The results are reported in {Table}~\ref{tab:GCE}. Note that GCE-GNN requires {loading} the global graph into video memory, and such information for the Yoochoose dataset is too large to fit in. Therefore, we make a compromise on the batch size for experiments on the Yoochoose dataset, decreasing it from 100 to 64. To make fair comparisons, we also rerun the experiments for SR-GNN and our G$^3$SR methods so the results reported here {are} slightly different from {Table~\ref{tab:comp}\footnote{That is the reason why we report these comparison results in appendix.}}. {Comparing to GC-SAN and SR-GNN, the superiority of introducing global information is obvious. Overall, combining G$^3$SR leads to the best results.}

  \bibliographystyle{IEEEtran}
  \bibliography{ref}

\vskip -0.3in
\begin{IEEEbiography}
[{\includegraphics[width=1in,height=1.25in,clip,keepaspectratio]{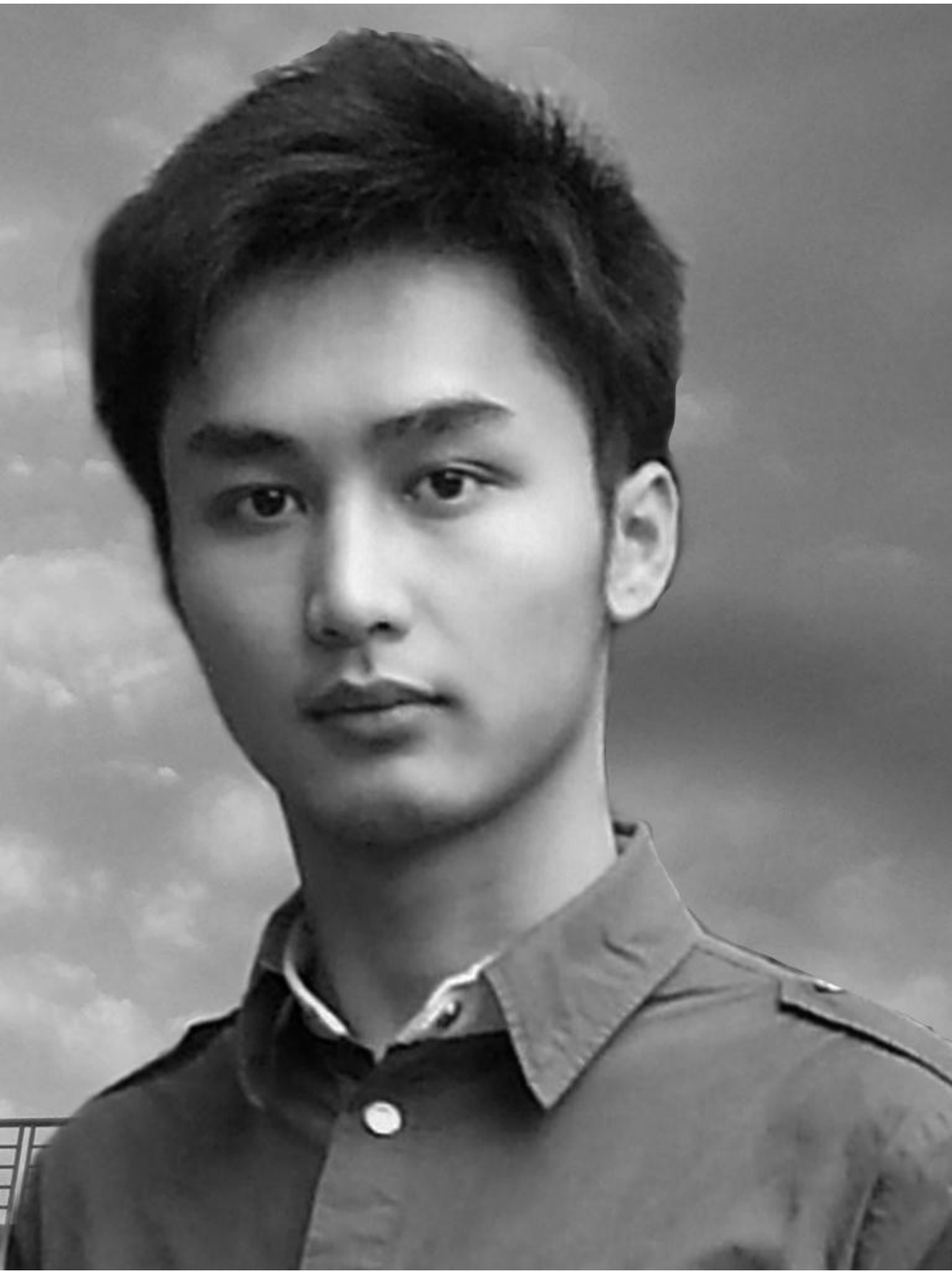}}]
{Zhi-Hong Deng} received his undergraduate and master degree in computer science in 2017 and 2020 respectively from Sun Yat-sen University. {He is pursuing a PhD degree at University of Technology Sydney. His research interests span data mining and machine learning, with a special focus on recommender systems and reinforcement learning.} He has published three papers in AAAI, ICLR and IEEE TCYB.
\end{IEEEbiography}

\vskip -0.4in
\begin{IEEEbiography}
[{\includegraphics[width=1in,height=1.25in,clip,keepaspectratio]{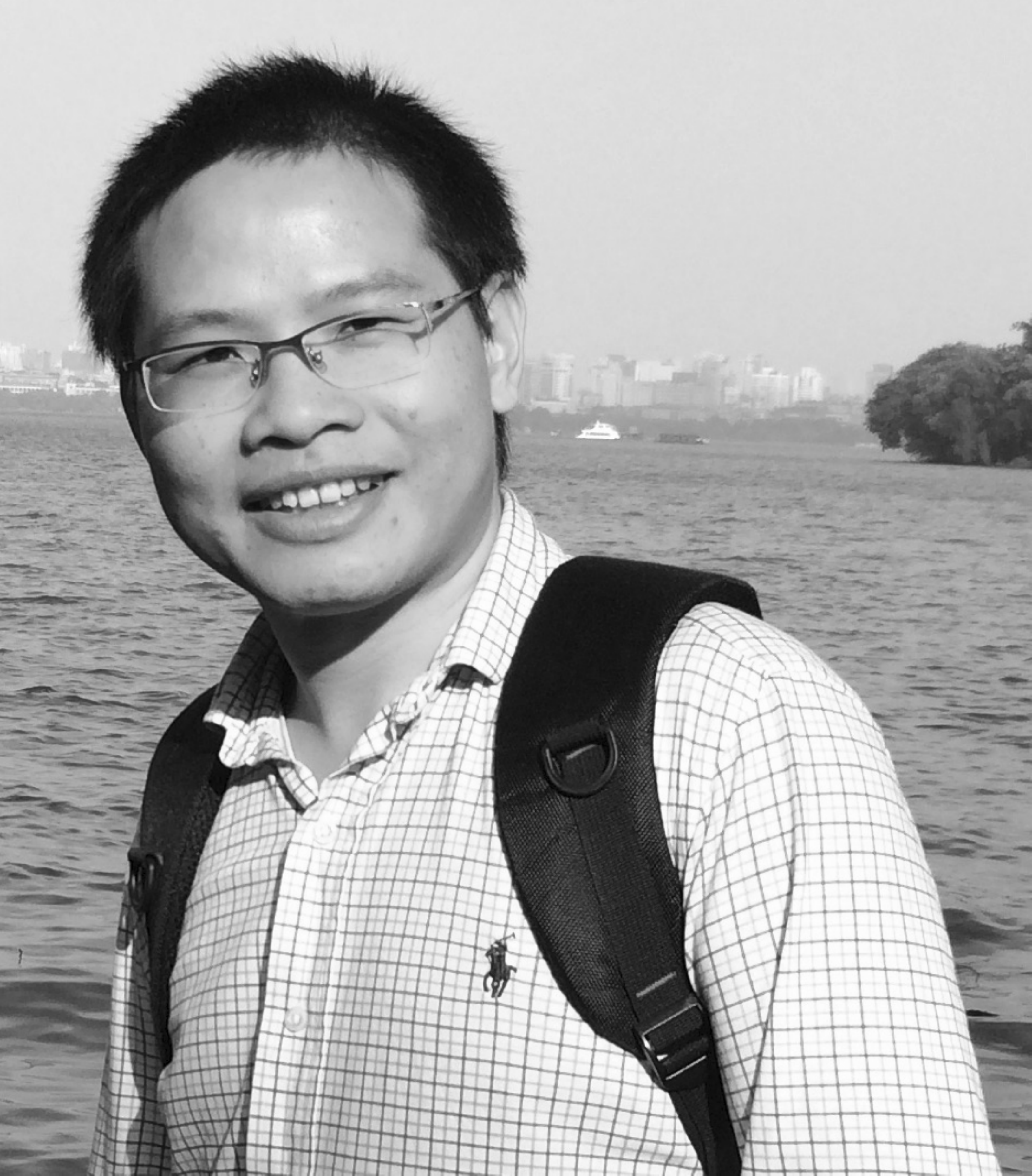}}]{Chang-Dong
Wang} received the Ph.D. degree in computer
science in 2013 from Sun Yat-sen University, Guangzhou,
China. He is a visiting student at University of Illinois at Chicago
from Jan. 2012 to Nov. 2012.
He joined Sun Yat-sen University in
2013, where he is currently an associate professor with School of Computer Science and Engineering.
His current research interests include machine learning
and data mining. He has published over 70 scientific papers in
international journals and conferences such as IEEE TPAMI, IEEE
TKDE, IEEE TCYB, IEEE TNNLS, ACM TKDD, IEEE TSMC-Systems, IEEE TII, IEEE TSMC-C, KDD, AAAI, IJCAI, CVPR, ICDM, CIKM and SDM. His ICDM 2010 paper won the Honorable Mention for Best
Research Paper Awards. He won 2012 Microsoft Research Fellowship Nomination Award. He was awarded 2015 Chinese Association for Artificial Intelligence (CAAI) Outstanding Dissertation. He is an Associate Editor in Journal of Artificial Intelligence Research (JAIR).
\\
\end{IEEEbiography}

% \vskip -0.7in
\begin{IEEEbiography}
[{\includegraphics[width=1in,height=1.25in,clip,keepaspectratio]{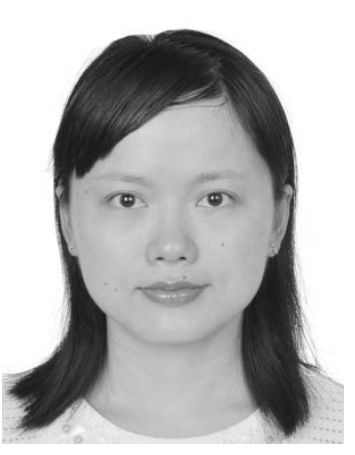}}]{Ling Huang} received her Ph.D. degree in computer science in 2020 from Sun Yat-sen University, Guangzhou. She joined South China Agricultural University in 2020 as an associate professor. She has published over 10 papers in international journals and conferences such as IEEE TKDE, IEEE TCYB, IEEE TNNLS, IEEE TII, ACM TKDD, IEEE/ACM TCBB, Pattern Recognition, KDD, AAAI, IJCAI and ICDM. Her research interest is data mining.
\end{IEEEbiography}

% \vskip -0.7in
\begin{IEEEbiography}[{\includegraphics[width=1in,height=1.25in,clip,keepaspectratio]{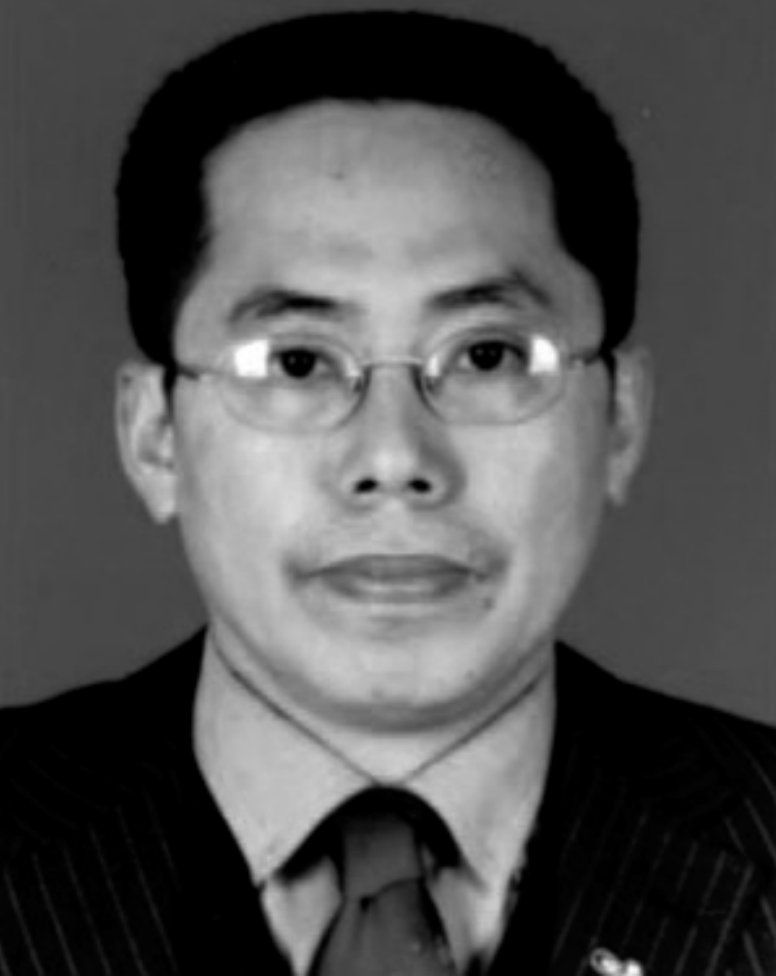}}]{Jian-Huang Lai}
received the M.Sc. degree in applied mathematics in 1989 and the Ph.D. degree in mathematics in 1999 from Sun Yat-sen University, China. He joined Sun Yat-sen University in 1989 as an Assistant Professor, where he is currently a Professor with the School of Data and Computer Science. His current research interests include the areas of digital image processing, pattern recognition, multimedia communication, wavelet and its applications. He has published more than 300 scientific papers in the international journals and conferences on image processing and pattern recognition, such as IEEE TPAMI, IEEE TKDE, IEEE TNNLS, IEEE TIP, ICCV, CVPR, IJCAI. Prof. Lai serves as a Standing Member of the Image and Graphics Association of China, and also serves as a Standing Director of the Image and Graphics Association of Guangdong.
\end{IEEEbiography}

% \vskip -0.7in
\begin{IEEEbiography}
[{\includegraphics[width=1in,height=1.25in,clip,keepaspectratio]{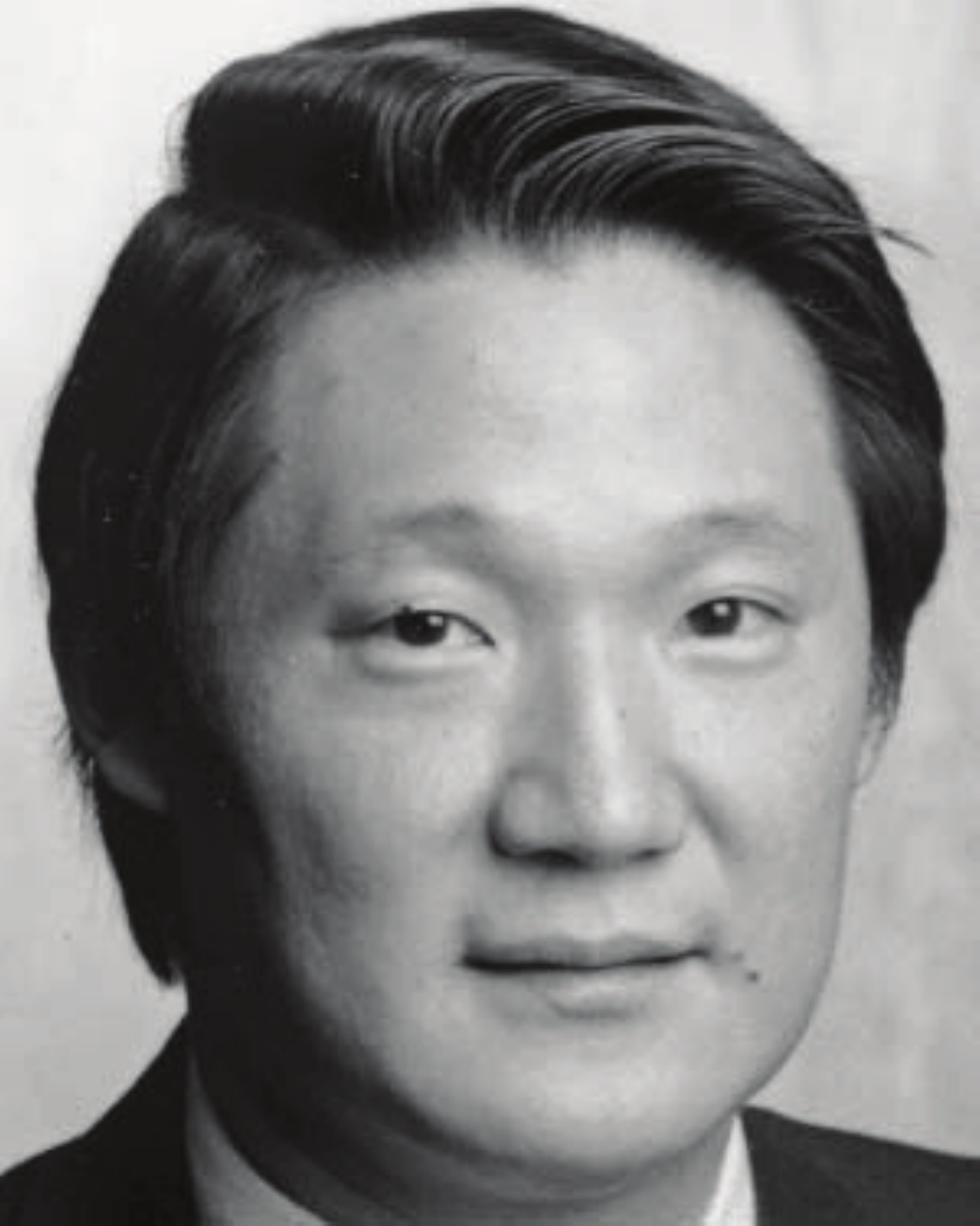}}]{Philip
S. Yu} is a Distinguished Professor in Computer Science at the University of Illinois at Chicago and also holds the Wexler Chair in Information Technology. Before joining UIC, Dr. Yu was with IBM, where he was manager of the Software Tools and Techniques group at the Watson Research Center. His research interest is on big data, including data mining, data stream, database and privacy. He has published more than 1,000 papers in refereed journals and conferences. He holds or has applied for more than 300 US patents.

Dr. Yu is a Fellow of the ACM and the IEEE. He was on the steering committee of the ACM Conference on Information and Knowledge Management and was a member of the steering committee of IEEE Data Engineering and IEEE Conference on Data Mining.  He was the Editor-in-Chief of ACM Transactions on Knowledge Discovery from Data (2011-2017) and the Editor-in-Chief of IEEE Transactions on Knowledge and Data Engineering (2001-2004).  Dr. Yu is the recipient of ACM SIGKDD 2016 Innovation Award for his influential research and scientific contributions on mining, fusion and anonymization of big data, the IEEE Computer Society's 2013 Technical Achievement Award for ``\textit{pioneering and fundamentally innovative contributions to the scalable indexing, querying, searching, mining and anonymization of big data}'', and the Research Contributions Award from IEEE Intl. Conference on Data Mining (ICDM) in 2003 for his pioneering contributions to the field of data mining. He also received the ICDM 2013 10-year Highest-Impact Paper Award, and the EDBT Test of Time Award (2014). He had received several IBM honors including 2 IBM Outstanding Innovation Awards, an Outstanding Technical Achievement Award, 2 Research Division Awards and the 94th plateau of Invention Achievement Awards.  He was an IBM Master Inventor. Dr. Yu received the B.S. Degree in E.E. from National Taiwan University, the M.S. and Ph.D. degrees in E.E. from Stanford University, and the M.B.A. degree from New York University.
\end{IEEEbiography}

 \end{document}